\documentclass[12pt,preprint]{aastex}

\shorttitle{Global 3D MHD Simulations of Galactic Disks}
\shortauthors{Nishikori et al.}

\begin{document}



\title{GLOBAL THREE-DIMENSIONAL MHD SIMULATIONS OF GALACTIC GASEOUS DISKS: \\ 
       I. AMPLIFICATION OF MEAN MAGNETIC FIELDS IN 
          AXISYMMETRIC GRAVITATIONAL POTENTIAL}


\author{Hiromitsu Nishikori\altaffilmark{1}}
\affil{Graduate School of Science and Technology, Chiba University, 
       1-33, Yayoi-cho, Inage-ku, Chiba 263-8522, Japan}
\email{nisikori@astro.s.chiba-u.ac.jp}

\author{Mami Machida\altaffilmark{2}}
\affil{National Astronomical Observatory of Japan, 
       2-21-1, Osawa, Mitaka, Tokyo 181-8588, Japan}

\and

\author{Ryoji Matsumoto\altaffilmark{3}}
\affil{Department of Physics, Faculty of Science, Chiba University, 
       1-33, Yayoi-cho, Inage-ku, Chiba 263-8522, Japan}




\begin{abstract}

We carried out global three-dimensional resistive magnetohydrodynamic 
simulations 
of galactic gaseous disks to investigate how the galactic magnetic fields are 
amplified and maintained.
We adopt a steady axisymmetric gravitational potential given 
by \citet*{Miy75} and \cite*{Miy80}. 
As the initial condition, we assume a warm ($T\sim10^{5}\,\, \mathrm{K}$) rotating gas 
torus centered at $\varpi=10\,\, \mathrm{kpc}$ threaded by weak azimuthal magnetic fields. 
Numerical results indicate that in differentially rotating galactic gaseous disks, 
magnetic fields are amplified due to magneto-rotational instability and 
magnetic turbulence develops. 
After the amplification of magnetic energy saturates,
the disk stays in a quasi-steady state. 
The mean azimuthal magnetic field increases with time and shows reversals 
with period of 1Gyr (2Gyr for a full cycle). 
The amplitude of $B_{\varphi}$ near the equatorial plane is 
$B_{\varphi}\sim1.5\mu\,\, \mathrm{G}$ at $\varpi=5\,\, \mathrm{kpc}$. 
The magnetic fields show large fluctuations whose standard deviation is 
comparable to the mean field.
The mean azimuthal magnetic field in the disk corona has direction opposite to the 
mean magnetic field inside the disk. 
The mass accretion rate driven by the Maxwell stress is $\sim10^{-3}M_{\odot}/\,\, \mathrm{yr}$
at $\varpi=2.5\,\, \mathrm{kpc}$ when the mass of the initial torus is 
$\sim5\times10^{8}M_{\odot}$.
When we adopt an absorbing boundary condition at $r=0.8\,\, \mathrm{kpc}$, the rotation curve 
obtained by numerical simulations almost coincides with the rotation curve of the stars 
and the dark matter. 
Thus even when magnetic fields are not negligible for gas dynamics of a 
spiral galaxy, 
galactic gravitational potential can be derived 
from observation of 
rotation curves using gas component of the disk. 

\end{abstract}


\keywords{galaxies: spiral---ISM: magnetic fields---MHD}


\section{INTRODUCTION}

Spiral galaxies have magnetic fields. 
In our Galaxy, the average magnetic field strength is a few $\mu$G. 
\citep*[e.g.,][]{Sof86,Bec96}. 
These magnetic fields have been explained by the dynamo action operating in 
galactic gas disks \citep*[e.g.,][]{Par71}.
In conventional theory of galactic dynamos, the following induction equation 
is solved; 
\begin{equation}
   \frac{\partial\bar{\mbox{\boldmath $B$}}}{\partial t} =
   \nabla\times(\mbox{\boldmath $v$}\times\bar{\mbox{\boldmath $B$}})
    +\eta \nabla^{2} \bar{\mbox{\boldmath $B$}} + 
     \nabla\times(\alpha \bar{\mbox{\boldmath$B$}}),
   \label{eqn1}
\end{equation}
where $\eta$ denotes the turbulent magnetic diffusivity, \mbox{\boldmath $v$} is the 
mean velocity, and \mbox{\boldmath $\bar{B}$} is the mean magnetic field. 
The last term on the right hand side of equation 
(\ref{eqn1}) is the dynamo term which represents the re-generation of mean 
magnetic fields ($\alpha$ effect). 
When velocity fields \mbox{\boldmath $v$} are given, equation (\ref{eqn1}) is 
linear with \mbox{\boldmath $\bar{B}$}. 
It can be solved with appropriate boundary conditions. 
This approach is referred to as the {\it kinematic dynamo}. 
The effects of nonlinearities such as the quenching of the 
$\alpha$-effect for strong magnetic fields and the loss of magnetic flux due to the 
magnetic buoyancy have also been studied in the framework of the kinematic dynamo 
theory \citep*[e.g.,][]{Sch89,Bra89,Bra92}.

In 1990s, it became clear that nonlinear dynamical processes are essential 
for the evolution of magnetic fields in differentially rotating disks. 
\citet*{Bal91} showed that in the presence of weak magnetic fields, 
differentially rotating gas disks subject to the robust magneto-rotational 
instability (MRI). 
This instability is caused by the back-reaction of magnetic fields to the 
fluid motion. 
Thus, in order to study the evolution of magnetic fields in differentially 
rotating disks, we should solve the momentum equation including Lorentz force 
simultaneously with the induction equation. 
This approach is called {\it dynamical dynamo}. 
The developments of computational magnetohydrodynamics (MHD) enabled us to 
study this dynamical dynamo process by direct three-dimensional (3D) 
global simulations. 

In the context of accretion disks, the nonlinear growth of MRI was first studied 
by local 3D MHD simulations \citep*[e.g.,][]{Haw95,Mat95,Bra95}.
They showed that MRI drives magnetic turbulence in accretion disks and amplifies 
magnetic fields. 
Subsequently, global 3D MHD simulations of initially weakly magnetized rotating 
torus were carried out \citep*[e.g.,][]{Mat99,Haw00,Mac00}.
They showed that the magnetic field amplification saturates when 
$\beta=P_{gas}/P_{mag} \sim 10$. 
The magnetic fields maintained in the quasi-steady state are mainly toroidal 
but have turbulent poloidal components. 
They also showed that in the innermost region of black hole accretion disks 
where radial infall of disk gas becomes important, spiral magnetic fields 
which accompany radial field reversals are created \citep*[e.g.,][]{Mac03}.

In numerical studies of accretion disks, the gravitational field is assumed 
to be that of the central point mass. 
It is straightforward to extend this gravitational potential to more general 
potentials such as those of spiral galaxies. 
Since the galactic gaseous disk is also a differentially rotating disk, 
it subjects to the MRI. 
The growth of MRI and generation of MHD turbulence in galactic HI disks were 
discussed by \citet*{Sel99}. 
\citet*{Kim03} carried out three-dimensional local MHD
simulations of galactic gaseous disks and showed that MRI drives
interstellar turbulence and helps forming the giant molecular clouds. 
The coupling of the thermal instability and MRI in the interstellar space has been 
studied by \citet*{Pio04}.

\citet*{Dzi03} reported the results of global three-dimensional 
MHD simulations of the galactic gaseous disks including the dynamo 
$\alpha$ term. 
More recently, \citet*{Dzi04} carried out global three-dimensional 
MHD simulations of galactic gaseous disks 
without assuming the dynamo $\alpha$ term 
starting from vertical magnetic fields 
and showed that mean magnetic fields and interstellar turbulence are generated. 
However, their simulation area was limited to 5kpc in the radial
direction and 1kpc in the vertical direction.
\citet*{Kit04} carried out linear but
global analysis for MRI in a disk geometry and showed that MRI can
amplify very weak seed magnetic fields in proto-galaxies. 

In this paper, we report the results of global 3D MHD simulations of galactic 
gaseous disks by using the axisymmetric gravitational potential created 
by stars and the dark matter. 
We solve the nonlinear MHD equations without introducing the phenomenological 
dynamo $\alpha$ parameter. 
The initial magnetic field is assumed to be toroidal. 
This initial condition enables us to start the simulation 
without much disturbing the interface between the disk and the 
disk corona. 
When we start from vertical magnetic fields, efficient angular momentum loss 
near the surface of the initial torus drives an avalanche-like surface accretion, 
which deforms magnetic field lines into an hourglass shape and produces 
magnetocentrifugally driven outflows \citep*[e.g.,][]{Mat96}. 
For simplicity we neglect the multi-phase nature 
of galactic gaseous disk, non-axisymmetric spiral gravitational potential, 
self gravity and supernova explosions. 
These processes will be incorporated in subsequent papers. 

In section 2, we present numerical methods. 
Numerical results are given in section 3. 
Section 4 is devoted for summary and discussion.

\section{NUMERICAL METHODS}

\subsection{Basic Equations}

The basic equations we solve are resistive MHD equations, 

\begin{equation}
   \frac{\partial \rho}{\partial t} + \nabla \cdot 
   (\rho \mbox{\boldmath $v$})=0, \label{eqn:eqn2.1}
\end{equation}
\begin{equation}
   \rho\left[\frac{\partial\mbox{\boldmath $v$}}{\partial t} + 
   (\mbox{\boldmath $v$}\cdot\nabla)\mbox{\boldmath $v$}\right] =
   - \nabla P + \frac{1}{4\pi}(\nabla \times \mbox{\boldmath $B$}) \times 
     \mbox{\boldmath $B$}
   - \rho \nabla \phi,
   \label{eqn:eqn2.2}
\end{equation}

\begin{equation}
   \frac{\partial\mbox{\boldmath $B$}}{\partial t} =
   \nabla\times(\mbox{\boldmath $v$}\times\mbox{\boldmath $B$} 
    -\eta \nabla \times \mbox{\boldmath $B$}),
   \label{eqn:eqn2.3}
\end{equation}

\begin{eqnarray}
   \frac{\partial}{\partial t}
   \left(\frac{1}{2}\rho v^2 + \frac{B^2}{8\pi} + \frac{P}{\gamma -1}\right) 
   \hspace{35mm} \nonumber\\ 
   + \nabla\cdot\left[\left(\frac{1}{2}\rho\mbox{\boldmath $v$}^2 +  
   \frac{\gamma P}{\gamma -1}\right)\mbox{\boldmath $v$}  + 
   \frac{c}{4\pi}\mbox{\boldmath $E$}\times\mbox{\boldmath $B$}\right]  
   \nonumber\\ 
   =  - \rho \mbox{\boldmath $v$} \nabla \phi .
   \label{eqn:eqn2.4}
\end{eqnarray}
In these equations, $\rho$, $p$, $\mbox{\boldmath $B$}$, $\mbox{\boldmath $v$}$, 
$\phi$ and $\gamma$ are the density, pressure, magnetic fields, 
velocity, gravitational potential and the specific heat ratio, respectively. 
The electric field \mbox{\boldmath $E$} is related to the magnetic field 
\mbox{\boldmath $B$} 
by 
\mbox{\boldmath $E$} = 
($-$\mbox{\boldmath $v$}$\times$\mbox{\boldmath $B$} + 
$\eta$$\nabla$$\times$\mbox{\boldmath $B$})/$c$.
We assume the anomalous resistivity $\eta$ 
\[
\eta = \left\{ 
\begin{array}{ll}
   \eta_{0} (v_{d}/v_{c}-1)^2, & (\mbox{when}\: v_{d} > v_{c}), \\
   0,                         & (\mbox{when}\: v_{d} < v_{c}), 
\end{array}
\right.
\]
where $v_{d}=j / \rho$ is the electron-ion drift velocity and 
$v_{c}$ is the critical velocity above which anomalous resistivity sets 
in \citep*{Yok94}.

We adopted cylindrical coordinates $(\varpi, \varphi, z)$. 
For normalization, we take the unit radius $\varpi_{0} = 1 \,\, \mathrm{kpc}$, 
and unit velocity $v_{0} = 255 \,\, \mathrm{km/s}$. 
The unit time is $t_{0} = \varpi_{0} / v_{0} = 3.8 \times 10^{6}$ years. 
For specific heat ratio and resistivity, we adopt $\gamma = 5 / 3$ and 
$\eta_{0}=0.01 v_{0}\varpi_{0}$, respectively.

\subsection{Initial Condition}

As the initial density distribution, we assume a rotating equilibrium torus 
threaded by azimuthal magnetic fields embedded 
in the isothermal, non-rotating, spherical hot halo \citep*{Oka89}. 
We assume that the gas torus has constant specific angular momentum $L$ 
and assume polytropic equation of state, $P = K \rho^{\gamma}$ where
$K$ is a constant. 
This initial condition simulates the situation that after a spiral 
galaxy is formed, the interstellar gas is supplied either by accretion of intergalactic 
gas 
which has some specific angular momentum or by supernova explosions following bursts of 
star formation inside the galaxy at some radius.
As we shall show later, the angular momentum distribution of the gas settles into that 
determined by the galactic gravitational potential. 
Thus, the final distribution of density and angular momentum is 
not sensitive to the initial distribution. 

We assume that the Alfv\'en speed $v_{A}$ is a function of $\varpi B_{\varphi}$, 
\begin{equation}
   v_{A}^{2} \equiv \frac{B_{\varphi}^2}{4\pi\rho} = 
   \frac{(4 \pi \mathcal{H})^{1/\mu}}{4\pi}
        (\varpi B_{\varphi})^{2(\mu-1)/\mu}, 
   \label{eqn:Alfven}
\end{equation}
where $B_{\varphi}$ is the toroidal magnetic field and $\mathcal{H}$ and
$\mu$ are constants.  
We take $\mu = \gamma$. 
Using these assumptions, we can integrate the equation of motion into the 
potential form, 
\begin{eqnarray}
   \Psi (\varpi , z) = \phi + \frac{L^2}{2\varpi^2} +
          \frac{1}{\gamma -1}v_{s}^{2} +
          \frac{\gamma}{2(\gamma-1)}v_{A}^{2}   
     = \Psi_{b} \nonumber \\ 
     = constant, 
   \label{eqn:pot1}
\end{eqnarray}
where $v_{s} = (\gamma P / \rho)^{1/2}$ 
is the sound speed and $\Psi_{b} = \Psi (\varpi_{b},0)$. 
We take the reference radius $\varpi_{b}$ at the radius
of the density maximum of the torus. 
We adopt $\varpi_{b} = 10 \,\, \mathrm{kpc}$ and set $\rho(\varpi_{b},0) = \rho_{b}$. 
Using equation (\ref{eqn:pot1}), we obtain the density distribution 

\begin{equation}
   \rho = \rho_{b}\left[
          \frac{\,\, \mathrm{max} \{ \Psi_{b} -\phi (\varpi,z) - L^{2}/(2\varpi^{2}),0 \} }
         {K\{\gamma/(\gamma -1)\}\left( 1+\beta_{b}^{-1}\varpi^{2(\gamma - 1)}
         /\varpi_{b}^{2(\gamma - 1)}\right)}
         \right]^{1/(\gamma - 1)}, 
   \label{eqn:density}
\end{equation}
where $\beta_{b} = (2K/\mathcal{H}) / \varpi_{b}^{2(\gamma - 1)}$ is the
ratio of gas pressure to magnetic pressure at $(\varpi,z) = (\varpi_{b},0)$. 
The thermal energy of the torus is parameterized by 
\begin{equation}
   E_{th}  =\frac{c_{sb}^{2}}{\gamma v_{0}^{2}}
	  =0.05,
   \label{eqn:ethermal}
\end{equation}
where 
$c_{sb}$ is the sound speed at $(\varpi, z) = (\varpi_{b}, 0)$. 
This sound speed corresponds to the temperature $T_{b} \sim 2 \times 10^{5} \,\, \mathrm{K}$. 
This means that we assume a warm torus as the initial condition. 
For unit density, we take 
$\rho_{b} = 3 \times 10^{-25} \,\, \mathrm{g/cm}^{3}$. 
The initial gas pressure at $(\varpi, z) = (\varpi_{b}, 0)$ is 
\begin{eqnarray}
   P_{b} &=& P(\varpi_{b},0) = \frac{1}{\gamma}\rho_{b}c_{sb}^{2}
   \nonumber \\
   &=& 10^{-11}\left(\frac{\rho_{b}}{3 \times 10^{-25}\,\, \mathrm{g/cm^{3}}}\right)
                   \left(\frac{T_{b}}{2 \times 10^{5}\,\, \mathrm{K}}\right)
   \,\, \mathrm{dyne/cm^{2}}. 
      \label{eqn:pressure}
\end{eqnarray}

The density of the corona $\rho_{c}$ is given by 
\begin{equation}
   \rho_{c} 
   =10^{-4}\rho_{b}\,\, \mathrm{exp}\left[ -\frac{\gamma}{c_{sc}^{2}}
     (\phi - \phi_{b})\right],
   \label{eqn:halo}
\end{equation}
where $c_{sc}$ is the sound speed in the corona and 
$\phi_{b}$ is the gravitational potential at $(\varpi, z) = (\varpi_{b}, 0)$.
We take the characteristic coronal pressure 
$P_{c} = \rho_{c}c_{sc}^{2} \sim 10^{-4}\rho_{b}c_{sc}^{2} = 3 \times 10^{-3}P_{b}$. 

As the gravitational potential $\phi$, we adopted a model of our galaxy
given by \citet*{Miy75} and \citet*{Miy80}. 
They assumed centrifugal balance only in the galactic plane. 
To reproduce the galactic rotation curve, three different sets of parameters 
$(a_{1}, b_{1}, M_{1})$, $(a_{2}, b_{2}, M_{2})$, $(a_{3}, b_{3}, M_{3})$ are 
introduced in the gravitational potential as follows, 
\begin{equation}
   \phi (\varpi , z) =
      \sum_{i=1}^{3}
        \frac{GM_{i}}{[\varpi^2 +
        \{a_{i} + (z^{2} + b_{i}^{2})^{0.5}\}^{2}]^{0.5}}.
   \label{eqn:MNpotential}
\end{equation}
Here $i = 1$ and $2$ take into account the gravitational field by bulge and 
disk stars, respectively. 
The third component $i = 3$ is the contribution by dark matter. 
In this equation, the constants $a_{i}$, $b_{i}$ and $M_{i}$ have dimension
of length and mass, respectively, and $G$ is the gravitational constant.
This potential is axisymmetric. 
The values of the constants are given in Table \ref{tbl:tbl1}. 
Model I and II are models without dark matter and model III-VII include
the dark matter. 

\begin{figure}[h]
\begin{center}
\epsscale{0.7}
\plotone{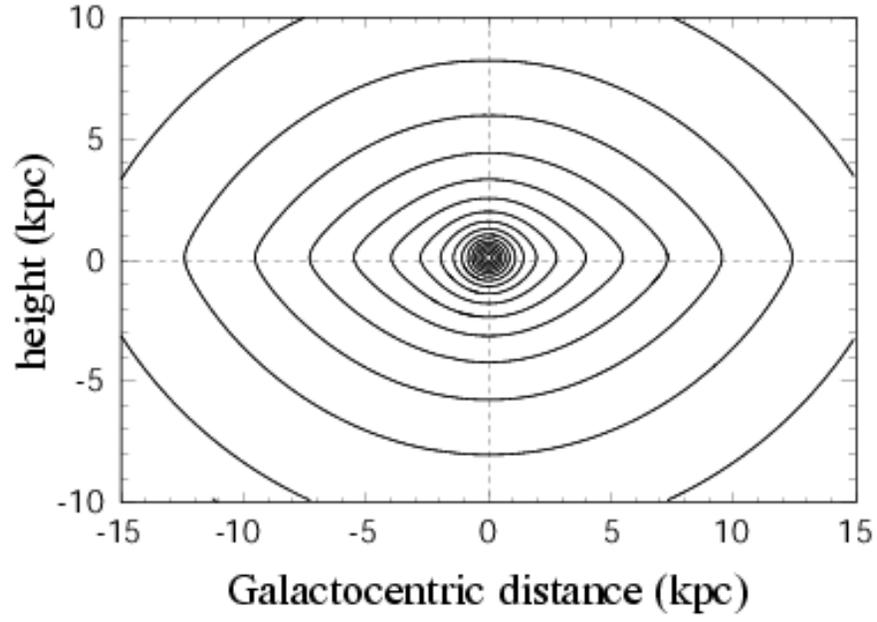}
\end{center}
\caption{Isocontours of the gravitational potential in 
         Miyamoto \& Nagai's (1975) model.
         This potential for model I does not include the effects of
         dark matter.
        }
\label{fig:f1}
\end{figure}

\begin{figure}[ht]
\begin{center}
\epsscale{0.6}
\plotone{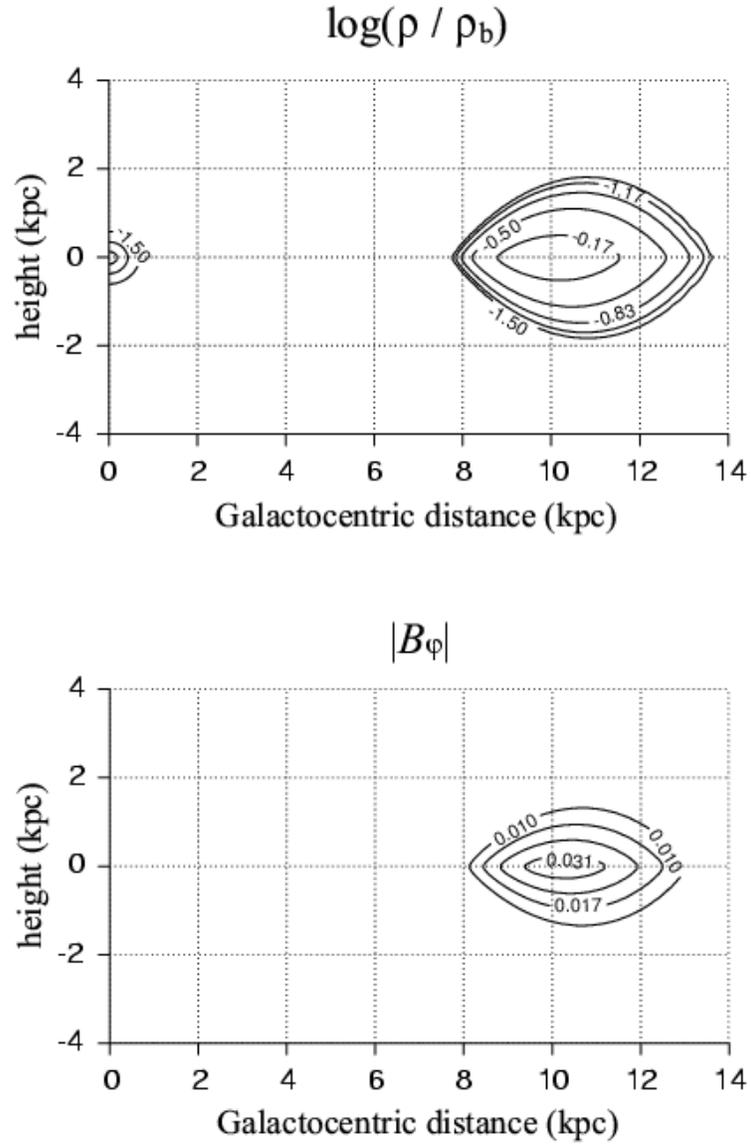}
\end{center}
\caption{Isocontours of initial density distribution
         (upper panel) and 
         azimuthal magnetic field (bottom panel) for model I. 
        }
\label{fig:f2}
\end{figure}

Figure \ref{fig:f1} shows the isocontours of the gravitational 
potential for model I. 
Figure \ref{fig:f2} shows the isocontours of the initial density and toroidal 
magnetic field for model I.

\subsection{Simulation Code}

We solved the MHD equations by using a three-dimensional 
MHD code based on a modified Lax-Wendroff method \citep*{Rub67}
with artificial viscosity \citep*{Ric67}. 
This code was originally developed to carry out
two-dimensional axisymmetric global MHD simulations of jet formation
from accretion disks \citep*[e.g.,][]{Shi85,Mat96,Hay96}. 
The MHD code was extended to three-dimensions and has been 
used to carry out global three-dimensional MHD simulations of accretion
disks (\citealt{Mac00}; \citealt*{Mac03,Kat04}).

\subsection{Simulation Region and Boundary Conditions}

The number of mesh points is $(N_{\varpi}, N_{\varphi}, N_{z}) = (250, 64, 319)$
for model I-IV.
The grid size is $\Delta \varpi = 0.05, \Delta z = 0.01$ for
$0 \leq \varpi < 6.0$, $0 \leq z < 2.0$, and otherwise 
increases with $\varpi$ and $z$ as $\Delta z_{k+1}= 1.05 \Delta z_{k}$. 
We used this nonuniform mesh to concentrate meshes near the galactic plane 
and near the central region. 
The size of the simulation box is $0 \,\, \mathrm{kpc} < \varpi < 56 \,\, \mathrm{kpc}$,
and $0 \,\, \mathrm{kpc} < z < 10 \,\, \mathrm{kpc}$. 
We used this large simulation box in order to 
simulate the buoyant rise of magnetic flux into the galactic corona and to 
reduce the effects of the outer boundary condition. 

We imposed free boundary conditions at $\varpi = 56 \,\, \mathrm{kpc}$ and 
$z = 10 \,\, \mathrm{kpc}$ where waves can be transmitted.
For azimuthal direction, we include full circle 
($0 \leq \varphi < 2 \pi$) for models I-IV.
We carried out full circle simulations because we are 
interested in the formation of large-scale mean magnetic fields. 
We should note that the amplitude of modes with low azimuthal mode numbers 
($m = 1$ or $m = 2$) 
are often large in global simulations of accretion disks
\citep*[e.g.,][]{Mac03}. 
In order to study the effects of the azimuthal resolution and 
the azimuthal size of the simulation region, 
we carried out simulations for model V, VI and VII, in which 
the azimuthal simulation region is limited to 0 $\leq \varphi < \pi/2$. 
Periodic boundary conditions are imposed in the azimuthal direction. 
The other parameters are the same as those for model III (Table \ref{tbl:tbl2}).

We assume symmetric boundary condition at the equatorial plane $z = 0$ and
simulated only the upper half space. 
The physical quantities at rotation axis ($\varpi = 0$) are computed by averaging 
those at the mesh point next to the axis. 
In models I and III-VII, we imposed absorbing boundary condition at 
$r = (\varpi^{2}+z^{2})^{1/2} = r_{in} = 0.8 \,\, \mathrm{kpc}$. 
We introduce a damping layer inside $r = r_{in}$. 
In this layer, the deviation of physical quantity $q$ from initial value $q_{0}$ is 
reduced after each time step with a damping rate $a$, 
   \begin{equation}
      q'=q-a(q-q_{0}).
   \end{equation}
Here we take 
   \begin{equation}
      a =
      0.1 \left( 1.0- \tanh \frac{r-0.8+5 \Delta \varpi}{2 \Delta \varpi } \right).
   \end{equation}
This damping layer serves as the non-reflecting boundary which
absorbs accreting mass and waves propagating into $r = 0.8$.
This boundary condition is adopted in our standard model because the gas 
accreted to the central region of the galaxy will be converted to stars or absorbed by 
the central massive black hole. 
In model II, we imposed no specific condition at $r = r_{in}$. 
Thus the accreting mass piles up in the central region.

\subsection{Initial Perturbation}

To initiate the evolution, we imposed random perturbation for the 
azimuthal velocity with maximum amplitude $0.01 v_{0}$. 
We checked the stability of the initial state by carrying out numerical simulations 
without including magnetic fields and without imposing velocity perturbations. 
Since the boundary between the rotating disk and the static corona 
is not exactly in hydrostatic equilibrium, small motions with maximum velocity 
$0.05 v_{0}$ appears in this region. 
The radial velocity in the dense region of the torus is less than 
$0.001 v_{0}$. 
The radial flow in the boundary layer between the disk and the corona slightly 
modifies the density distribution and smears the difference of the rotation speed 
between the disk and the corona. 
However, the flow does not significantly modify the density distribution of 
the main part of the torus. 
At $t = 2 \,\, \mathrm{Gyr}$, the maximum radial speed is about $0.0045 v_{0}$ 
even at the interface between the disk and the corona. 
The torus stays in the equilibrium state for time scale longer than 
$2.3 \,\, \mathrm{Gyrs}$.

\section{NUMERICAL RESULTS}

\subsection{Time Evolution of the Model with Absorbing  Inner Boundary Condition}

\begin{figure}
\epsscale{0.8}
\plotone{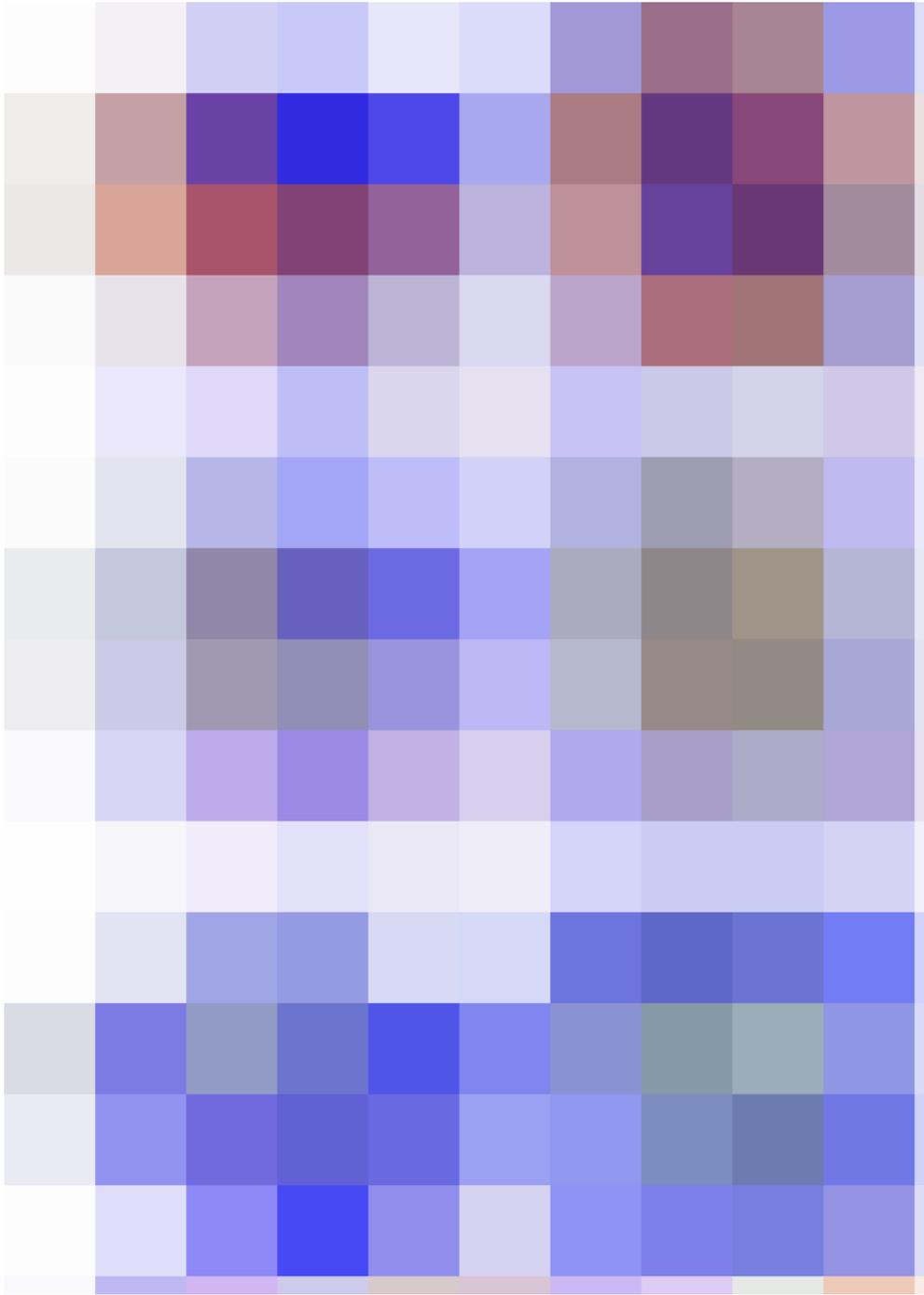}
\caption{Time evolution of density distribution 
         $\log(\rho / \rho_{b})$ (color scale) and magnetic field 
         lines (gray curves) in model I.
         The box size of left and right panels are
         $30 \,\, \mathrm{kpc} \times 30 \,\, \mathrm{kpc} \times 30 \,\, \mathrm{kpc}$
         and $30 \,\, \mathrm{kpc} \times 30 \,\, \mathrm{kpc}$, respectively.
	 The right panels show the equatorial density and magnetic field lines 
	 projected onto the equatorial plane.}
\label{fig:f3}
\end{figure}

First we present the results of Model I (the standard model). 
The inner boundary is treated as an absorbing boundary where accreting mass 
and magnetic fields are absorbed. 
Figure \ref{fig:f3} shows the time evolution of model I. 
Color contours show density distribution and solid curves show magnetic field lines. 
The left panels show the isosurface of the density
($\rho / \rho_{b} = 10^{-1.5}$), density distribution 
at the equatorial plane and magnetic field lines. 
We plotted the region $-15 \,\, \mathrm{kpc} < x, y, z < 15 \,\, \mathrm{kpc}$. 
The right panels show the equatorial density and magnetic field lines 
projected onto the equatorial plane. 
As the MRI grows, matter accretes to the central region by losing the 
angular momentum. 
The initial torus deforms itself into a flat disk. 
On the other hand, the matter in the outer part of the disk gets 
angular momentum and expands. 
The magnetic fields globally show spiral structure and locally show
turbulent structure. 

\begin{figure}[ht]
\begin{center}
\epsscale{0.5}
\plotone{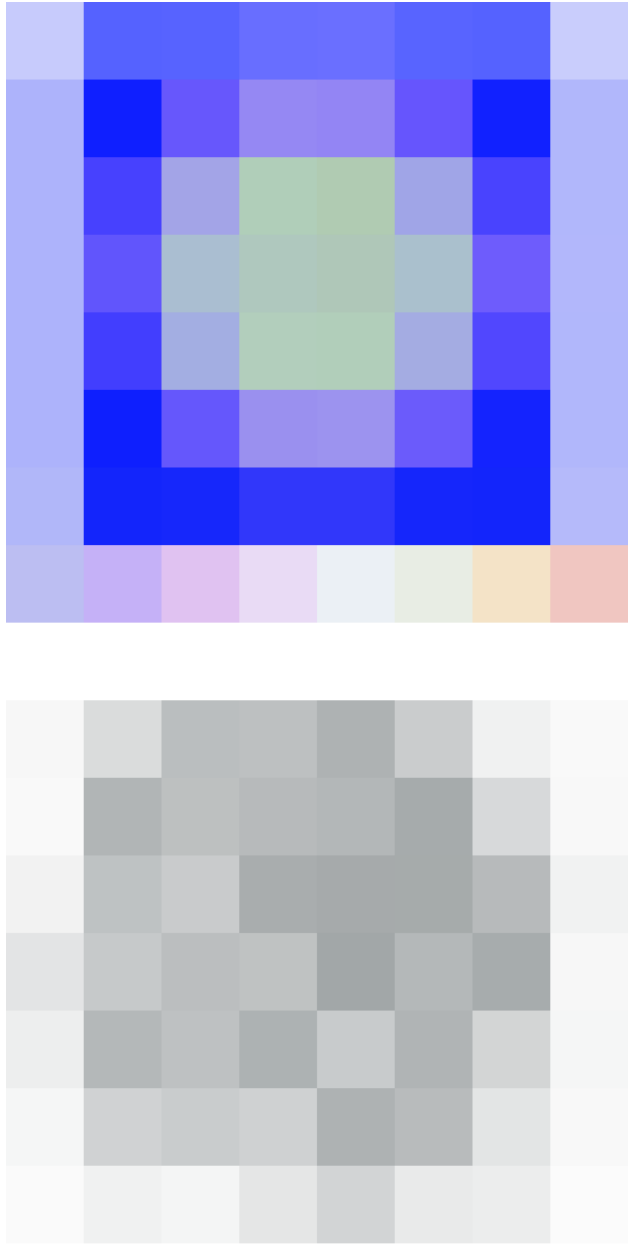}
\end{center}
\caption{Upper panel: 
         The distribution of density $\log(\rho / \rho_{b})$ for model I 
         in the equatorial plane at $t = 3.8 \,\, \mathrm{Gyr}$. 
         Lower panel: 
         magnetic field lines in model I
         projected onto the equatorial plane at $t = 3.8 \,\, \mathrm{Gyr}$.
         The box size is $30 \,\, \mathrm{kpc} \times 30 \,\, \mathrm{kpc}$. 
         }
\label{fig:f4}
\end{figure}

Figure \ref{fig:f4} shows the close up view of the density distribution 
and magnetic field lines projected onto the equatorial plane at 
$t = 1000 t_{0} (= 3.8 \,\, \mathrm{Gyr})$.
The box size is $-15 \,\, \mathrm{kpc} < x,y < 15 \,\, \mathrm{kpc}$.
The density distribution shows weak non-axisymmetric spiral pattern. 

\begin{figure}[p!]
\begin{center}
\epsscale{1.0}
\plotone{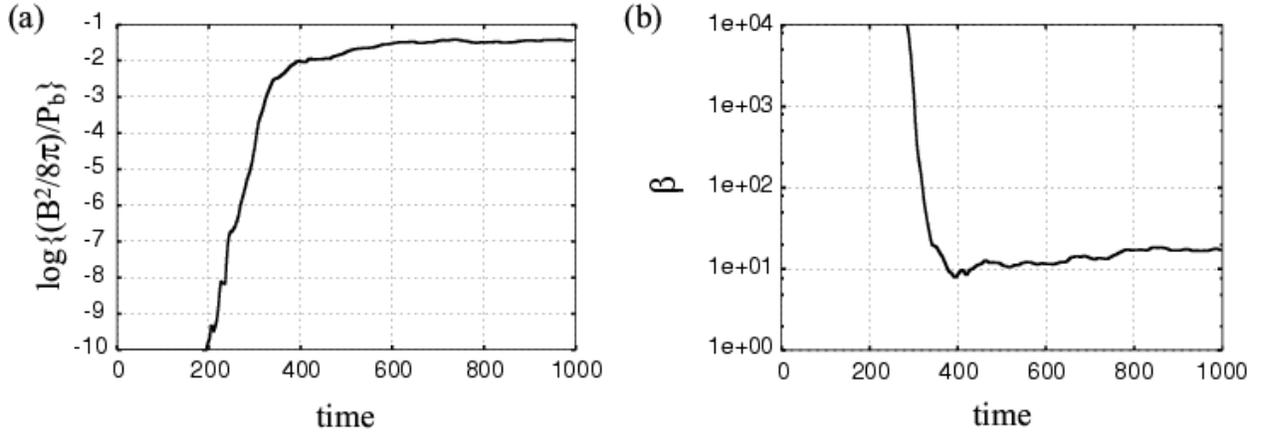}
\end{center}
\caption{(a)Time evolution of the magnetic energy and (b)volume averaged plasma 
         $\beta (= \langle P \rangle / \langle B^{2}/8\pi \rangle)$ 
         for model I averaged in $2 \,\, \mathrm{kpc} < \varpi < 5 \,\, \mathrm{kpc}$, 
         $0 \,\, \mathrm{kpc} < z < 1 \,\, \mathrm{kpc}$ and 
         $0 \leq \varphi < 2 \pi$.
	}
\label{fig:f5}
\end{figure}

Figure \ref{fig:f5}{\it a} shows the time evolution of magnetic energy 
for model I averaged in $2 \,\, \mathrm{kpc} < \varpi < 5 \,\, \mathrm{kpc}$, 
$0 \,\, \mathrm{kpc} < z < 1 \,\, \mathrm{kpc}$ and $0 \leq \varphi < 2\pi$. 
Figure \ref{fig:f5}{\it b} shows the plasma $\beta$ averaged in the same region. 
As the MRI grows in the disk, the magnetic energy increases and 
its strength saturates when $\log [(B^{2} / 8\pi) / P_{b}] \sim -1.5$. 
For our Galaxy, this saturation level corresponds to $B \sim 3 \mu \,\, \mathrm{G}$. 
The magnetic energy is maintained for time scale at least 
$2 \,\, \mathrm{Gyrs}$. 
As magnetic energy is amplified, the average value of 
plasma $\beta \equiv \langle P \rangle / \langle B^{2} / 8\pi \rangle$ 
decreases and stays around $\langle \beta \rangle \sim 20$ 
where $\langle \beta \rangle$ 
is the spatial average of local plasma $\beta$ 
in $2.0 \,\, \mathrm{kpc} < \varpi < 5.0 \,\, \mathrm{kpc}$, 
$0 \,\, \mathrm{kpc} < z < 1.0 \,\, \mathrm{kpc}$ and $0 \leq \varphi < 2 \pi$. 
This saturation level of magnetic energy is smaller than that observed in 
our galaxy. 
Further amplification of magnetic energy by, for example, supernova 
explosions may be necessary to explain the strength of magnetic fields in our galaxy. 

\begin{figure}[p!]
\begin{center}
\epsscale{1.0}
\plotone{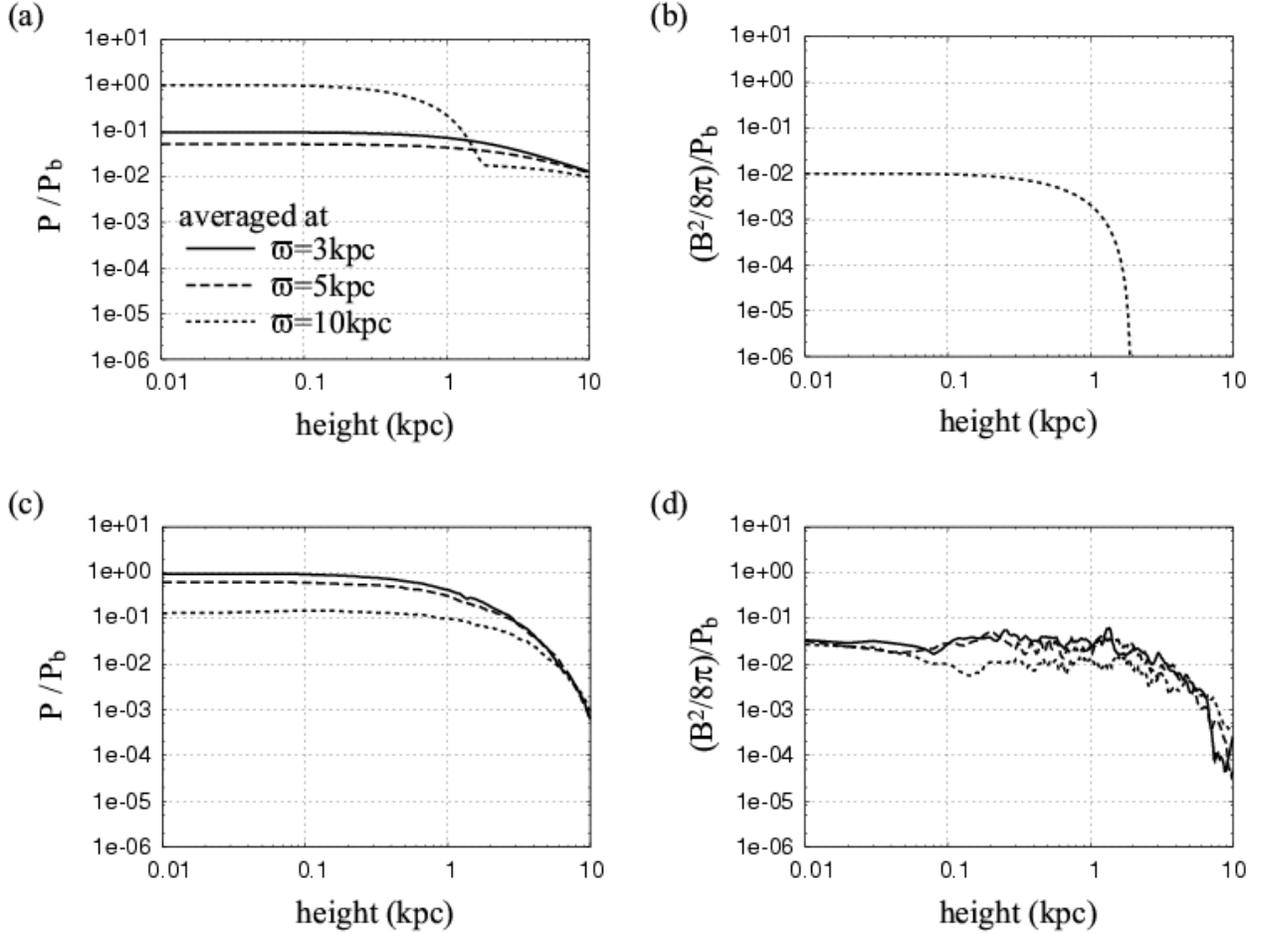}
\end{center}
\caption{Vertical distribution of the gas pressure and magnetic pressure. 
         (a) Initial distribution of gas pressure. 
	 (b) Initial distribution of magnetic pressure
         (c) vertical distribution of gas pressure at $t = 1000t_{0}$  and 
         (d) the vertical distribution of magnetic pressure at $t = 1000t_{0}$. 
         The solid, dashed and dotted curves indicate quantities averaged at 
         $\varpi = 3 \,\, \mathrm{kpc}$, $5 \,\, \mathrm{kpc}$ and $10 \,\, \mathrm{kpc}$, 
         respectively.
        } 
\label{fig:f6}
\end{figure}

Magnetic field lines shown in the left panels of figure
\ref{fig:f3} indicate that magnetic fields initially confined in the
disk emerge from the disk and form extended magnetized corona. 
Figure \ref{fig:f6} shows the vertical distribution of 
the gas pressure and the magnetic pressure at $\varpi = 3 \,\, \mathrm{kpc}$, 
$5 \,\, \mathrm{kpc}$ and $10 \,\, \mathrm{kpc}$ at $t = 0$ and
$t = 1000t_{0} ( = 3.8 \,\, \mathrm{Gyr})$. 
The magnetic flux initially confined in the torus fills almost the whole
simulation region. 
The plasma $\beta$ in the corona is $1 < \beta < 10$ in
$1 \,\, \mathrm{kpc} < z < 10 \,\, \mathrm{kpc}$.

Magnetic fields emerge from the disk to the corona due to the growth of
the Parker instability \citep*{Par66}.
Nonlinear growth of Parker instability in disks was studied by
\citet*{Mat88} by two-dimensional MHD simulations taking into account 
the vertical variation of the gravitational field. 
The effects of the differential 
rotation and MRI on the Parker instability
were studied by \citet*{Fog94,Fog95}. 
\citet*{Mil00} carried out local three-dimensional MHD
simulations of gravitationally stratified differentially rotating disks
and showed that strongly magnetized corona is created due to the buoyant
rise of magnetic flux from the disk to the corona. 
\citet{Mac00} showed by global three-dimensional MHD simulations
that buoyantly rising magnetic loops are formed above accretion disks. 
Our numerical results are consistent with these previous results.

\begin{figure}[p!]
\epsscale{1.0}
\plotone{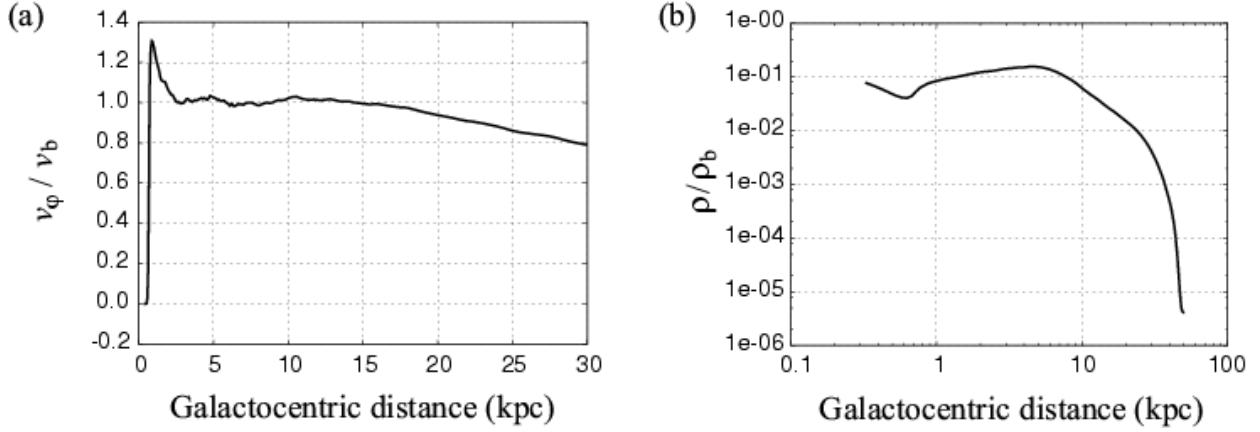}
\caption{The distribution of (a)azimuthal velocity and  
         (b)density $\rho$ at $t = 3.8 \,\, \mathrm{Gyr}$ averaged in 
         $0 \,\, \mathrm{kpc} < z < 0.3 \,\, \mathrm{kpc}$ 
	 and $0 \leq \varphi < 2 \pi$ for model I. 
	}
\label{fig:f7}
\end{figure}

Figure \ref{fig:f7} shows the radial distribution of azimuthal velocity 
$v_{\varphi}$ and the density $\rho$ averaged in 
$0 \,\, \mathrm{kpc} < z < 0.3 \,\, \mathrm{kpc}$ and 
$0 \leq \varphi < 2\pi$ at $t = 3.8 \,\, \mathrm{Gyr}$. 
In the inner region ($\varpi < 10 \,\, \mathrm{kpc}$), the radial profile of
azimuthal velocity of model I is similar to that observed in our Galaxy. 
The density distribution in Figure \ref{fig:f7}{\it b} indicates that the initial 
torus spreads and becomes flat in the inner region. 

\begin{figure}[p!]
\begin{center}
\epsscale{1.0}
\plotone{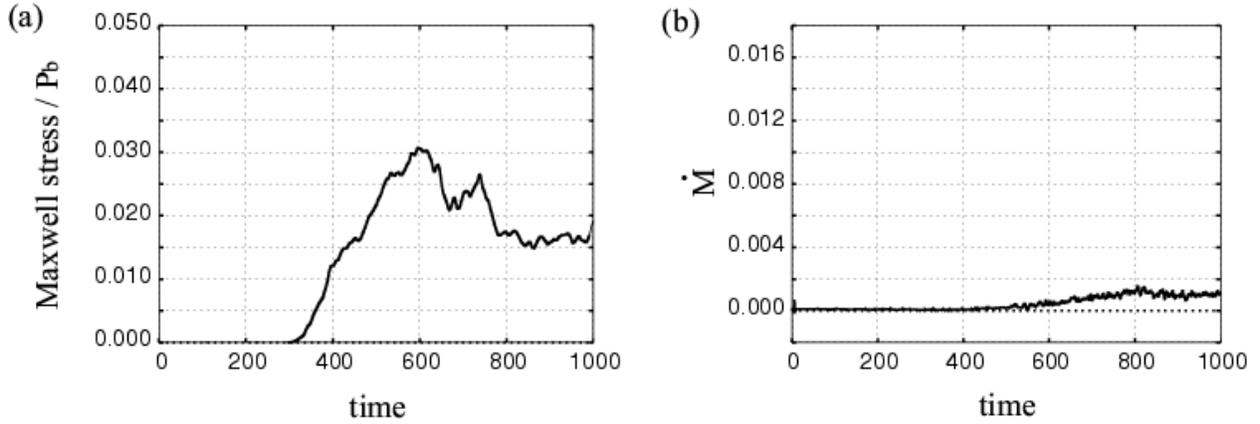}
\end{center}
\caption{(a)Time evolution of Maxwell stress normalized by $P_{b}$ for model I. 
         The Maxwell stress is averaged in 
	 $2 \,\, \mathrm{kpc} < \varpi < 5 \,\, \mathrm{kpc}$, 
         $0 \,\, \mathrm{kpc} < z < 1 \,\, \mathrm{kpc}$ and $0 \leq \varphi < 2 \pi$. 
         (b)Time evolution of the accretion rate at $2.5 \,\, \mathrm{kpc}$ from
         the center.
         The unit of the accretion rate is $\dot{M_{0}} = 1 M_{\odot}/ \,\, \mathrm{yr}$ 
         when $\rho_{b} = 3 \times 10^{-25} \,\, \mathrm{g/cm}^3$. 
        }
\label{fig:f8}
\end{figure}

Figure \ref{fig:f8}{\it a} shows the time evolution of 
the ratio of Maxwell stress averaged in 
$2 \,\, \mathrm{kpc} < \varpi < 5 \,\, \mathrm{kpc}$, 
$0 \,\, \mathrm{kpc} < z < 1 \,\, \mathrm{kpc}$ and 
$0 \leq \varphi < 2 \pi$ to the initial gas pressure at $\varpi = 10 \,\, \mathrm{kpc}$,
\begin{equation}
   \alpha_{B} \equiv \left\langle - \frac{B_{\varpi} B_{\varphi}}{4 \pi P_{b}}\right\rangle
   . 
\label{eqn:Maxwell_stress}
\end{equation}
The Maxwell stress increases as the gas accretes and 
later decreases gradually.

Figure \ref{fig:f8}{\it b} shows the time evolution of the
accretion rate at $\varpi = 2.5$kpc defined by
\begin{equation}
   \dot M_{\varpi=2.5} =
        - \int_{0}^{2\pi} \int_{0}^{2.5kpc} \rho v_{\varpi} \varpi d \varphi dz.
\label{eqn:M_dot}
\end{equation}
The unit of the accretion rate is $1 M_{\odot} / \,\, \mathrm{yr}$ when 
$\rho_{b} = 3 \times 10^{-25} \,\, \mathrm{g/cm}^{3}$. 
Numerical result indicates that $\dot M \sim 10^{-3}M_{\odot} / \,\, \mathrm{yr}$. 
Since the mass of the initial torus is $M_{torus} = 5 \times 10^{8}M_{\odot}$, 
mass accretion takes place for time scale longer than $10 \,\, \mathrm{Gyr}$.

\subsection{Effects of the Central Absorber and Dark Matter}

In this subsection, we show the effects of the central absorber 
and the dark matter. 
In model I, we assumed that the interstellar matter accreting to the central 
region are absorbed at $r = 0.8 \,\, \mathrm{kpc}$. 
This mimics the existence of the central black hole or the efficient conversion 
of gas to stars in the central region. 
When such absorber does not exist, accreting matter piles up in the central 
region. 
As a reference model, we carried out a simulation of model II, in which 
we allow the accreting mass to pile up in the central region. 

\begin{figure}[p!]
\begin{center}
\epsscale{0.5}
\plotone{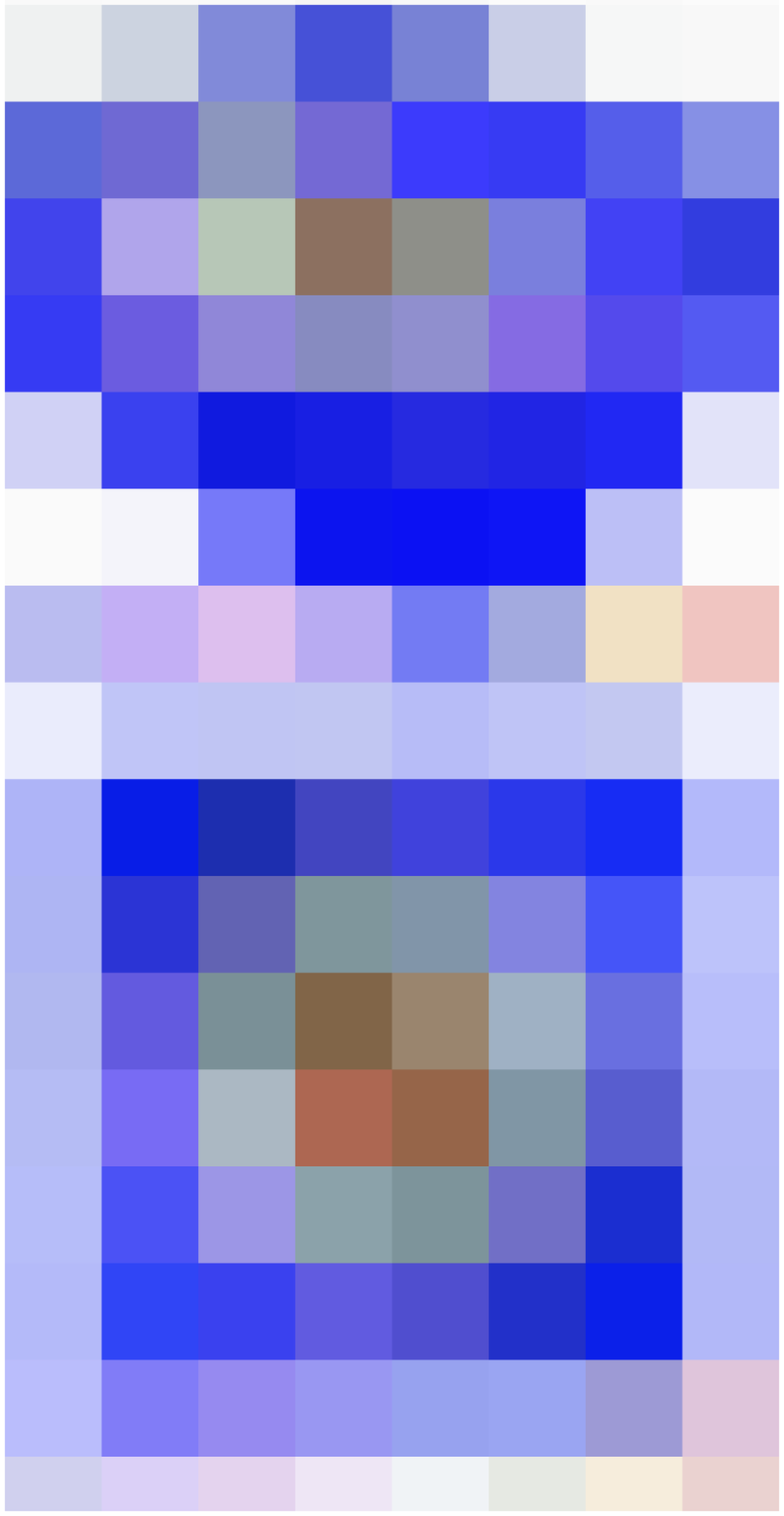}
\end{center}
\caption{The distribution of density (color scale) and magnetic field lines 
         (gray curves) at $t = 3.8 \,\, \mathrm{Gyr}$ for model II. 
	 The box size is $30 \,\, \mathrm{kpc} \times 30 \,\, \mathrm{kpc}$. 
        }
\label{fig:f9}
\end{figure}

Figure \ref{fig:f9} shows the density distribution and magnetic field
lines at $t = 1000t_{0} ( = 3.8 \,\, \mathrm{Gyr})$ for model II. 
Since accreted matter accumulates in the central region, 
it forms a high density bulge.

\begin{figure}[p!]
\begin{center}
\epsscale{0.5}
\plotone{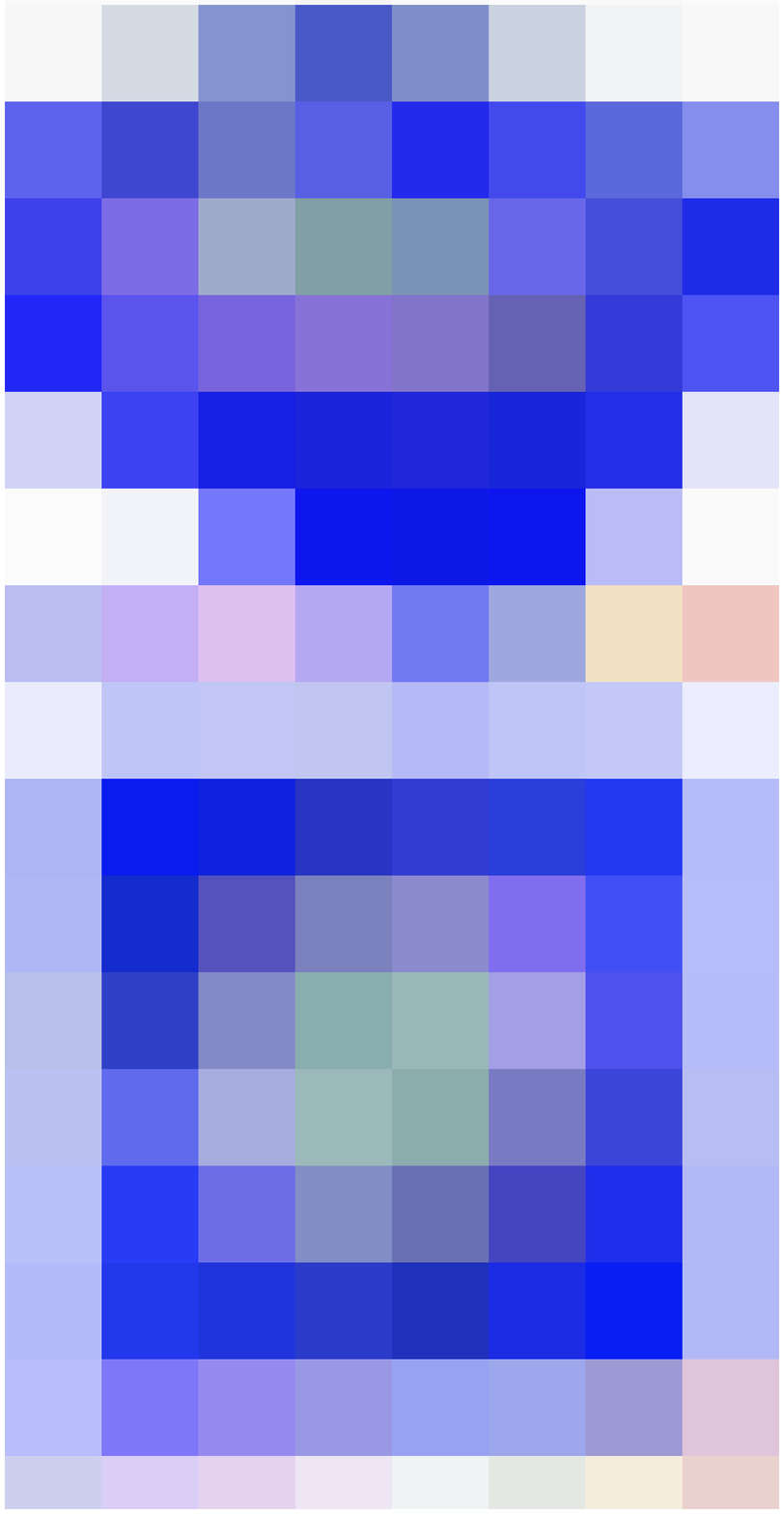}
\end{center}
\caption{The distribution of density (color scale) and magnetic field lines
         (gray curves) at $t = 3.8\,\, \mathrm{Gyr}$ for model III.
         The box size is $30 \,\, \mathrm{kpc} \times 30 \,\, \mathrm{kpc}$. 
        }
\label{fig:f10}
\end{figure}

Figure \ref{fig:f10} shows the distribution of density and magnetic field lines 
at $t = 3.8 \,\, \mathrm{Gyr}$ for model III. 
In this model, the effect of dark matter is included. 
Absorbing boundary condition is imposed at $r = 0.8 \,\, \mathrm{kpc}$.
The density distribution and magnetic field lines are similar to those of model I. 

\begin{figure}[p!]
\begin{center}
\epsscale{1.0}
\plotone{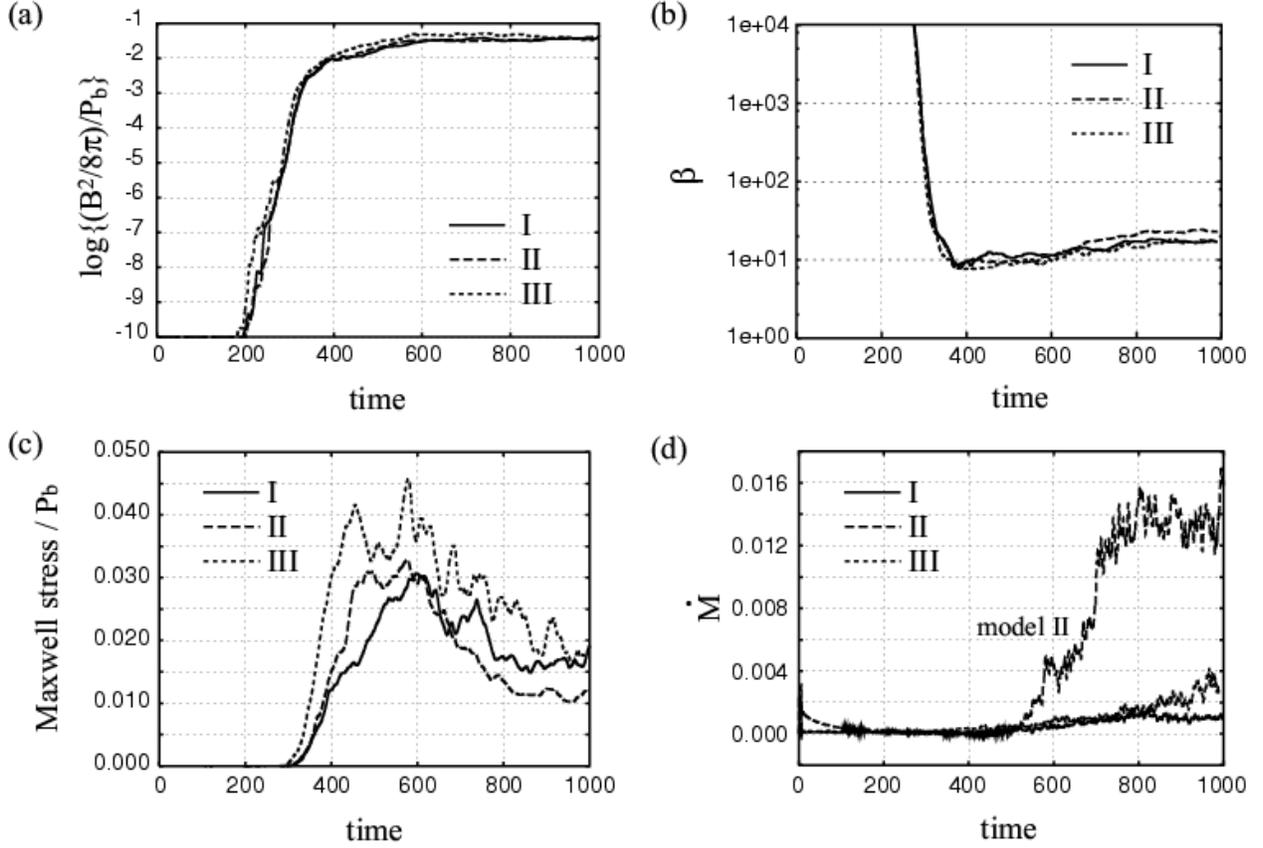}
\end{center}
\caption{Comparison of numerical results for model I, model II and model III. 
  	 (a)Time evolution of the magnetic energy. 
	 (b)Time evolution of the plasma $\beta$ averaged in 
	    $2 \,\, \mathrm{kpc} < \varpi < 5 \,\, \mathrm{kpc}$, 
            $0 \,\, \mathrm{kpc} < z < 1 \,\, \mathrm{kpc}$ and $0 \leq \varphi < 2 \pi$. 
         (c)Maxwell stress averaged in the same region.
	    The solid, dashed and dotted curves show the results for model I, 
	    model II and model III, respectively. 
	 (d)The time evolution of the accretion rate at $2.5 \,\, \mathrm{kpc}$ from the center.
         }
\label{fig:f11}
\end{figure}

\begin{figure}[p!]
\begin{center}
\epsscale{1.0}
\plotone{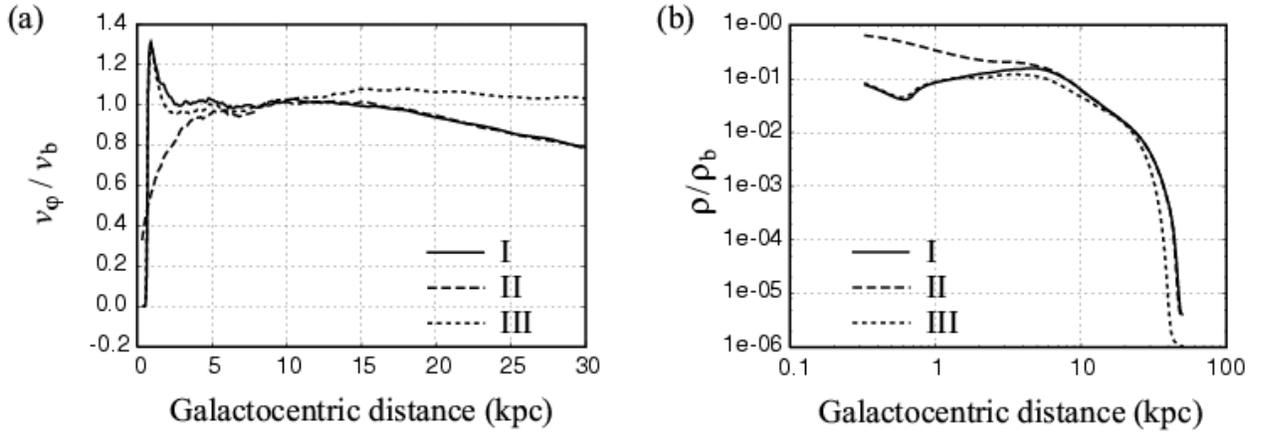}
\end{center}
\caption{(a)The radial distribution of azimuthal velocity $v_{\varphi}$ and 
         (b)radial distribution of density $\rho$ averaged in 
	 $0 \,\, \mathrm{kpc} < z < 0.3 \,\, \mathrm{kpc}$ and 
         $0 \leq \varphi < 2 \pi$ at $t = 3.8 \,\, \mathrm{Gyr}$. 
	 The solid, dashed and dotted curves show the results for
         model I, model II and model III, respectively.}
\label{fig:f12}
\end{figure}

Figures \ref{fig:f11}{\it a} and \ref{fig:f11}{\it b} show the time evolution of
magnetic energy and plasma $\beta$ averaged in 
$2 \,\, \mathrm{kpc} < \varpi < 5 \,\, \mathrm{kpc}$, 
$0 \,\, \mathrm{kpc} < z < 1 \,\, \mathrm{kpc}$ and $0 \leq \varphi < 2 \pi$, 
for models I, II and III. 
The magnetic energy saturates at approximately the same level for all models. 
The magnetic energy is maintained more than 2Gyrs.
However, after significant amount of gas accretes, the value of plasma $\beta$ 
in model II becomes larger than that of other models, because in model II, 
the gas density and pressure increase with time. 
Figure \ref{fig:f11}{\it c} shows the ratio of Maxwell stress to gas pressure
averaged in $2 \,\, \mathrm{kpc} < \varpi < 5 \,\, \mathrm{kpc}$, 
$0 \,\, \mathrm{kpc} < z < 1 \,\, \mathrm{kpc}$ and $0 \leq \varphi < 2 \pi$ 
for model I, II and III. 
The time evolution of Maxwell stress has almost the same profile in all models.
Figure \ref{fig:f11}{\it d} shows the time evolution of the accretion rate 
at $\varpi = 2.5 \,\, \mathrm{kpc}$ for these models. 
In model II, the accretion rate is about 10 times larger 
than that of model I and model III because the density near 
$\varpi = 2.5 \,\, \mathrm{kpc}$ of model II is larger than the density of
model I and model III (see figure \ref{fig:f12}{\it b}).

Figure \ref{fig:f12} shows the distribution of azimuthal velocity 
$v_{\varphi}$ and density $\rho$ averaged in 
$0 \,\, \mathrm{kpc} < z < 0.3 \,\, \mathrm{kpc}$ and 
$0 \leq \varphi < 2 \pi$ at $t = 3.8 \,\, \mathrm{Gyr}$. 
The azimuthal velocity of model I and model III is similar to the
observed profile of our Galaxy in the inner region. 
The velocity profile of model III coincides with 
the observed profile in all region because the effects of dark matter
are included. 
On the other hand, the velocity distribution of model II is different 
from the observed rotation curve. 
In the inner region, the density of model II is ten times larger than that of 
model I and model III because the accreted gas piles up in the central region. 
When mass piles up in the central region, since the
pressure gradient force in the radial direction reduces the effective
gravity in the radial direction, the equilibrium rotation speed becomes
much less than the observed rotation speed. 
Thus, the existence of the central absorber or conversion of infalling
gas to stars are essential to reproduce the observed rotation curve. 
In the outer region ($8 \,\, \mathrm{kpc} < \varpi$) velocity and density of
model I and model II almost coincide. 

\subsection{Spatial and Temporal Reversals of Mean Magnetic Fields}\label{sec:mmf}

Figure \ref{fig:f13} shows the magnetic field lines
depicted by mean magnetic fields for model III at $t = 3.8 \,\, \mathrm{Gyr}$. 
Regions colored in orange or blue show domains where mean azimuthal magnetic
field at $z = 0.25\,\, \mathrm{kpc}$ is positive or negative, respectively.
The box size is $30 \,\, \mathrm{kpc} \times 30 \,\, \mathrm{kpc}$. 
We computed the mean magnetic field at each grid point
\begin{equation}
   {\bar{\mbox{\boldmath $B$}}}(\varpi_{i},\varphi_{j},z_{k}) =
   \frac{
     \sum^{i+10}_{l=i-10} \sum^{j+3}_{m=j-3} \sum^{k+10}_{n=k-10} 
         \mbox{\boldmath $B$} (\varpi_{l},\varphi_{m},z_{n})
   }
        {21 \times 7 \times 21},
\end{equation}
by averaging magnetic fields inside $\pm 10$ meshes in $\varpi$ and 
$z$ directions, and $\pm 3$ meshes in $\varphi$ direction. 
The mean magnetic field is mainly toroidal but shows reversals of toroidal 
components around $\varpi \sim 5 \,\, \mathrm{kpc}$ at this time. 

The azimuthal direction of mean magnetic fields in the disk changes with
radius in regions where channel-like flow develops. 
\citet*{Haw92} showed by axisymmetric simulations that channel-like flows
appear in the nonlinear stage of MRI. 
Even when the unperturbed magnetic field is purely azimuthal, spiral
channel flows appear as a result of the nonlinear growth of the
non-axisymmetric MRI \citep*{Mac03}. 
Since such magnetic channels move inward as the
infalling gas loses angular momentum, the radius of the field reversal
changes with time.

\begin{figure}[p!]
\begin{center}
\epsscale{0.7}
\plotone{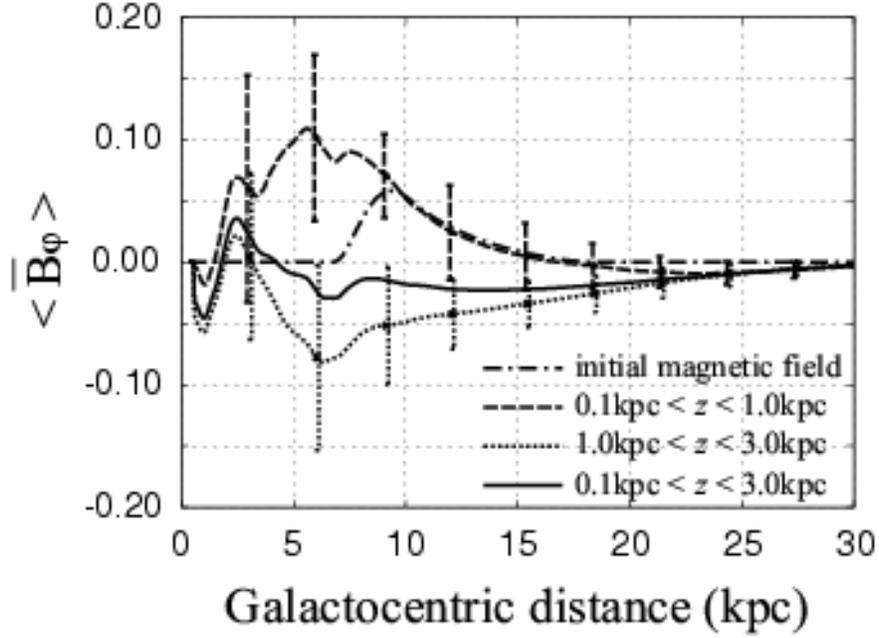}
\end{center}
\caption{Mean azimuthal field and its standard deviation (vertical bars) for model III 
         at $t = 3.1\,\, \mathrm{Gyr}$ averaged in azimuthal direction 
         ($0 \leq \varphi < 2 \pi$) and 
         in $0.1 \,\, \mathrm{kpc} < z < 1 \,\, \mathrm{kpc}$ (dashed curve), 
         $1.0 \,\, \mathrm{kpc} < z < 3.0 \,\, \mathrm{kpc}$ (dotted curve) and  
         $0.1 \,\, \mathrm{kpc} < z < 3.0 \,\, \mathrm{kpc}$ (solid curve), respectively.
         The unit of $B_{\varphi}$ is $\sqrt{\rho_{b}v_{0}^{2}} = 15 \mu \,\, \mathrm{G}$ 
         in this model when $\rho_{b} = 3 \times 10^{-25} \,\, \mathrm{g/cm}^{3}$.
        }
\label{fig:f14}
\end{figure}

Figure \ref{fig:f14} shows the azimuthally averaged mean toroidal field 
$\langle \bar{B_{\varphi}} \rangle$ (mean field) and the standard
deviation of azimuthal field 
$\sqrt{\langle (B_{\varphi}-{\bar{B_{\varphi}}})^{2}} \rangle$ 
(fluctuating field) at $t = 3.1\,\, \mathrm{Gyr}$ 
where $\langle$ $\rangle$ denotes the spatial average. 
The dashed, dotted and solid curves show the distribution of mean azimuthal 
magnetic fields averaged in azimuthal direction ($0 \leq \varphi < 2 \pi$) 
and in $0.1 \,\, \mathrm{kpc} < z < 1.0 \,\, \mathrm{kpc}$ (equatorial region), 
$1.0 \,\, \mathrm{kpc} < z < 3.0 \,\, \mathrm{kpc}$ (coronal region) and 
$0.1 \,\, \mathrm{kpc} < z < 3.0 \,\, \mathrm{kpc}$, respectively. 
The vertical bars show the standard deviation of the azimuthal field. 
The strength of fluctuating field is comparable to the mean magnetic field.
The dash-dotted curve shows the initial distribution of azimuthal mean magnetic 
fields averaged in 0.3kpc $< z <$ 2.0kpc and 0 $\leq \varphi < 2 \pi$. 
The strength of equatorial azimuthal magnetic field 
($0.1 \,\, \mathrm{kpc} < z < 1 \,\, \mathrm{kpc}$) is 
larger than the initial azimuthal field by factor 2. 

The mean azimuthal magnetic field in the coronal region 
($1.0 \,\, \mathrm{kpc} < z < 3.0 \,\, \mathrm{kpc}$) has direction opposite to
that in the disk region. 
The total magnetic flux threading the {\it $\varpi$-z}
plane is almost conserved. 

\begin{figure}[p!]
\begin{center}
\epsscale{0.5}
\plotone{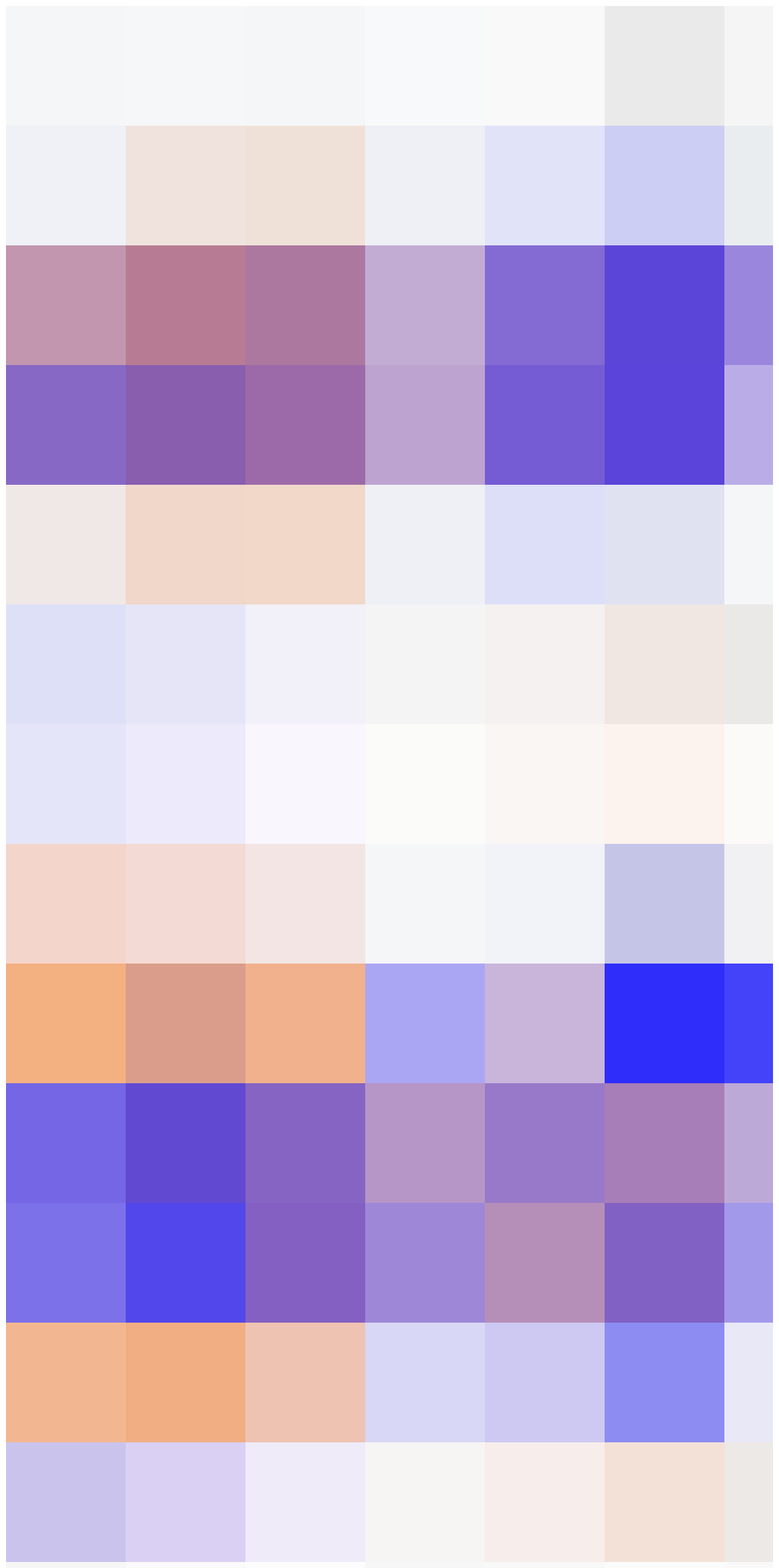}
\end{center}
\caption{The distribution of mean azimuthal magnetic fields for model
         III at $t=590t_{0}(= 2.2 \,\, \mathrm{Gyr})$ and $826 t_{0}( = 3.1 \,\, \mathrm{Gyr})$. 
         Blue and orange indicate regions where the mean field 
         $\bar{B_{y}}$ threading the $y=0$ plane is 
         positive or negative, respectively. 
         Arrows show the direction of mean magnetic fields. 
         The mean azimuthal fields change their direction with height 
         from the equatorial plane. 
        }
\label{fig:f15}
\end{figure}

Figure \ref{fig:f15} shows the spatial distribution of mean 
azimuthal magnetic fields for model III at $t = 590t_{0} (=2.2
\,\, \mathrm{Gyr})$ and $826t_{0} (= 3.1 \,\, \mathrm{Gyr})$. 
Blue and orange indicate regions where mean magnetic field 
$\bar{B_{y}}$ threading the $y = 0$ plane is 
positive or negative, respectively. 
Arrows show the direction of magnetic fields. 
These panels show that the mean azimuthal field
$\bar{B_{y}}$ threading the $y=0$ plane reverses its direction in the
coronal region. 
The maximum height of the magnetized region (the wavefront of the
azimuthal magnetic field) at $\varpi = 10 \,\, \mathrm{kpc}$ locates at $z =
4 \,\, \mathrm{kpc}$ at $t = 2.2 \,\, \mathrm{Gyr}$ and $z \sim 10 \,\, \mathrm{kpc}$
at $t = 3.1 \,\, \mathrm{Gyr}$. 
It indicates that the azimuthal magnetic flux is rising. 

The seeds for the reversals of the azimuthal magnetic fields 
are generated during the growth of MRI
because when the initial magnetic field is purely azimuthal, the growth
rate of the MRI is larger when $q = (k_{y}/k_{z})^{2}$ is small, where
$k_{y}$ and $k_{z}$ are the wavenumber in the azimuthal direction and
the vertical direction, respectively \citep*[e.g.,][]{Mat95}. 
Thus, the azimuthal direction of the perturbed magnetic field changes
with height. 
In the nonlinear stage when the vertical magnetic fields are produced by
the buoyant rise of the azimuthal magnetic flux, the formation of the
channel flows \citep*[e.g.,][]{Haw92} due to the nonlinear growth of MRI
also produces the reversals of azimuthal magnetic fields. 

\begin{figure}[p!]
\begin{center}
\epsscale{1.0}
\plotone{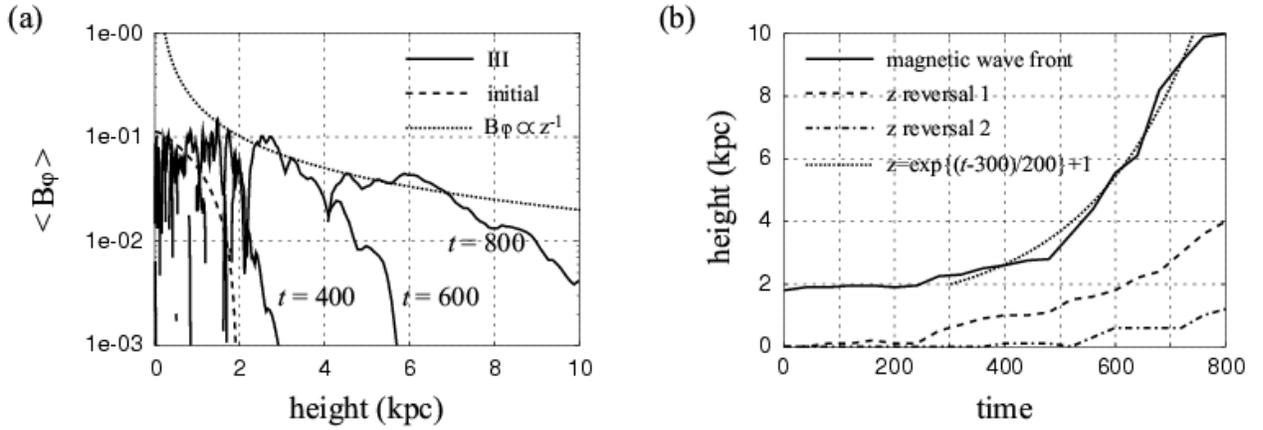}
\end{center}
\caption{(a) Vertical distribution of azimuthally averaged magnetic field
         $\langle B_{\varphi} \rangle$ at $\varpi = 10 \,\, \mathrm{kpc}$ 
         at $t = 400t_{0}$, $600t_{0}$ and $800t_{0}$ for model III. 
         The dashed curve shows the initial profile of $\langle B_{\varphi} \rangle$. 
         The dotted curve shows the $B_{\varphi} \propto z^{-1}$
         relation expected from the nonlinear theory of the Parker instability.
         (b) The time evolution of the wave front of rising magnetic
         flux where $B_{\varphi} = 0.003 \sqrt{\rho_{b} v_{0}^{2}}$ (solid
         curve) and the height of where direction of azimuthal magnetic field 
         reverses (dashed curve and dash-dotted curve) for model III. 
         The dotted curve shows the relation $z = 1 + \exp{\{(t-300)/200]\}}$
         expected from the nonlinear theory of the Parker instability.
        }
\label{fig:f16}
\end{figure}

Figure \ref{fig:f16}{\it a} shows the time variation of the vertical
distribution of azimuthally averaged magnetic field $\langle B_{\varphi}
\rangle$ for model III at $\varpi = 10 \,\, \mathrm{kpc}$. 
The regions where $\langle B_{\varphi} \rangle < 0$ are not plotted. 
The wavefront of the azimuthal flux propagates in the vertical direction
with speed $v \sim 8 \,\, \mathrm{kpc}/1.5 \,\, \mathrm{Gyr} \sim 5
\,\, \mathrm{km/s}$. 
This speed is comparable to the Alfv\'{e}n speed. 
The vertical distribution of the azimuthal field $\langle B_{\varphi} \rangle$
approximately follows the self-similar solution of the nonlinear Parker
instability $\langle B_{\varphi} \rangle \propto z^{-1}$ \citep{Shi89}. 
Figure \ref{fig:f16}{\it b} shows the time evolution of the height of
the wavefront of the azimuthal magnetic flux (solid curve) and the
height of reversals of azimuthal magnetic fields (dashed curve and
dash-dotted curve) at $\varpi = 10 \,\, \mathrm{kpc}$. 
The exponential increase of the height of the wavefront
is consistent with the self-similar solutions of the nonlinear Parker
instability \citep{Shi89}. 
After an azimuthal magnetic flux with one direction rises, another
azimuthal magnetic flux with opposite direction rises. 
The MRI unstable disk continuously feeds the corona with magnetic
fields. 

\begin{figure}[p!]
\begin{center}
\epsscale{0.7}
\plotone{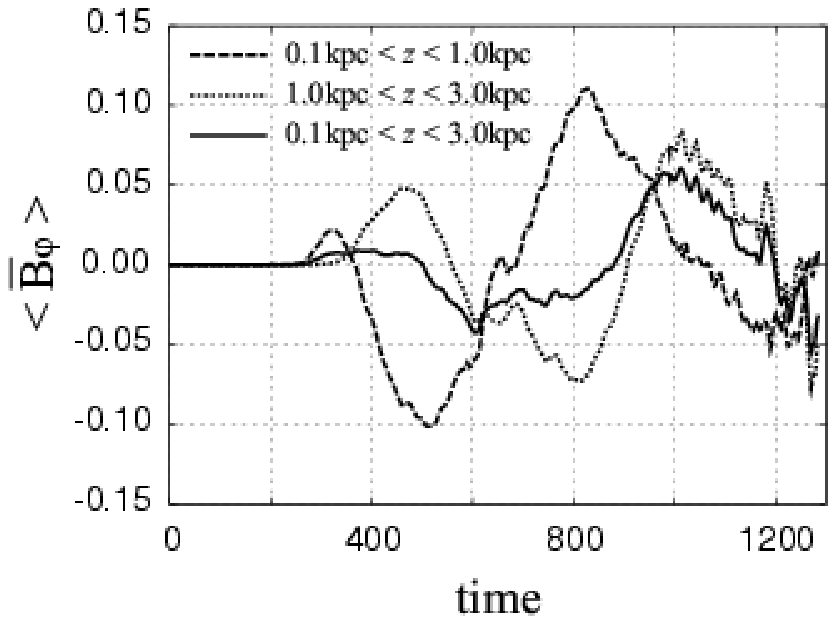}
\end{center}
\caption{The time variation of mean azimuthal field of model III 
         averaged in $5 \,\, \mathrm{kpc} < \varpi < 6 \,\, \mathrm{kpc}$, 
         $0 \leq \varphi < 2 \pi$ and
         $0.1 \,\, \mathrm{kpc} < z < 1 \,\, \mathrm{kpc}$ (dashed curve), 
         $1.0 \,\, \mathrm{kpc} < z < 3.0 \,\, \mathrm{kpc}$ (dotted curve) and
         $0.1 \,\, \mathrm{kpc} < z < 3.0 \,\, \mathrm{kpc}$ (solid curve), respectively.
         The unit of $B_{\varphi}$ is $\sqrt{\rho_{b}v_{0}^{2}} = 15 \mu \,\, \mathrm{G}$
         in this model when $\rho_{b} = 3 \times 10^{-25} \,\, \mathrm{g/cm}^{3}$.
        }
\label{fig:f17}
\end{figure}

In figure \ref{fig:f17} we plotted the time variation of mean azimuthal
field of model III. 
Magnetic fields are averaged in azimuthal direction 
($0 \leq \varphi <  2 \pi$), radial direction 
($5 \,\, \mathrm{kpc} \leq \varpi < 6 \,\, \mathrm{kpc}$), and 
in $0.1 \,\, \mathrm{kpc} < z < 1.0 \,\, \mathrm{kpc}$ (equatorial region), 
$1.0 \,\, \mathrm{kpc} < z < 3.0 \,\, \mathrm{kpc}$ (coronal region) and 
$0.1 \,\, \mathrm{kpc} < z < 3.0 \,\, \mathrm{kpc}$ (disk + corona), 
respectively. 
The equatorial magnetic field in $5\,\, \mathrm{kpc}<\varpi<6\,\, \mathrm{kpc}$ 
changes direction with time. 
The azimuthal magnetic fields reverse their direction with timescale 
$t \sim 300t_{0} (\sim 1 \,\, \mathrm{Gyr})$. 
This timescale is comparable to that of the buoyant rise of azimuthal
magnetic flux, which takes place in $t \sim 10 H/v_{A}$, 
\citep[e.g.,][]{Par66}, where $H$ is the half thickness of the disk 
($\sim 0.5 \,\, \mathrm{kpc}$), and $v_{A}$ is the Alfv\'{e}n speed around
the disk-corona interface ($v_{A} \sim 5 \,\, \mathrm{km/s}$). 

Here we would like to point out that the amplification of 
equatorial magnetic field is enabled by the escape of magnetic flux 
from the disk to the corona.
When the total azimuthal magnetic flux is conserved, the azimuthal
magnetic flux inside the disk can increase when magnetic flux escapes
from the disk to the corona. 
Although azimuthal magnetic flux is not exactly conserved 
in resistive MHD simulations, the flux is nearly conserved because
$\eta$ is small. 
Thus the buoyant escape of azimuthal magnetic flux from the disk to the
corona is compensated by the amplification of azimuthal magnetic flux
inside the disk with polarity opposite to that in the corona.

\subsection{Dependence on the Strength of Initial Magnetic Fields}
\label{sec:dep}

\begin{figure}[p!]
\begin{center}
\epsscale{0.5}
\plotone{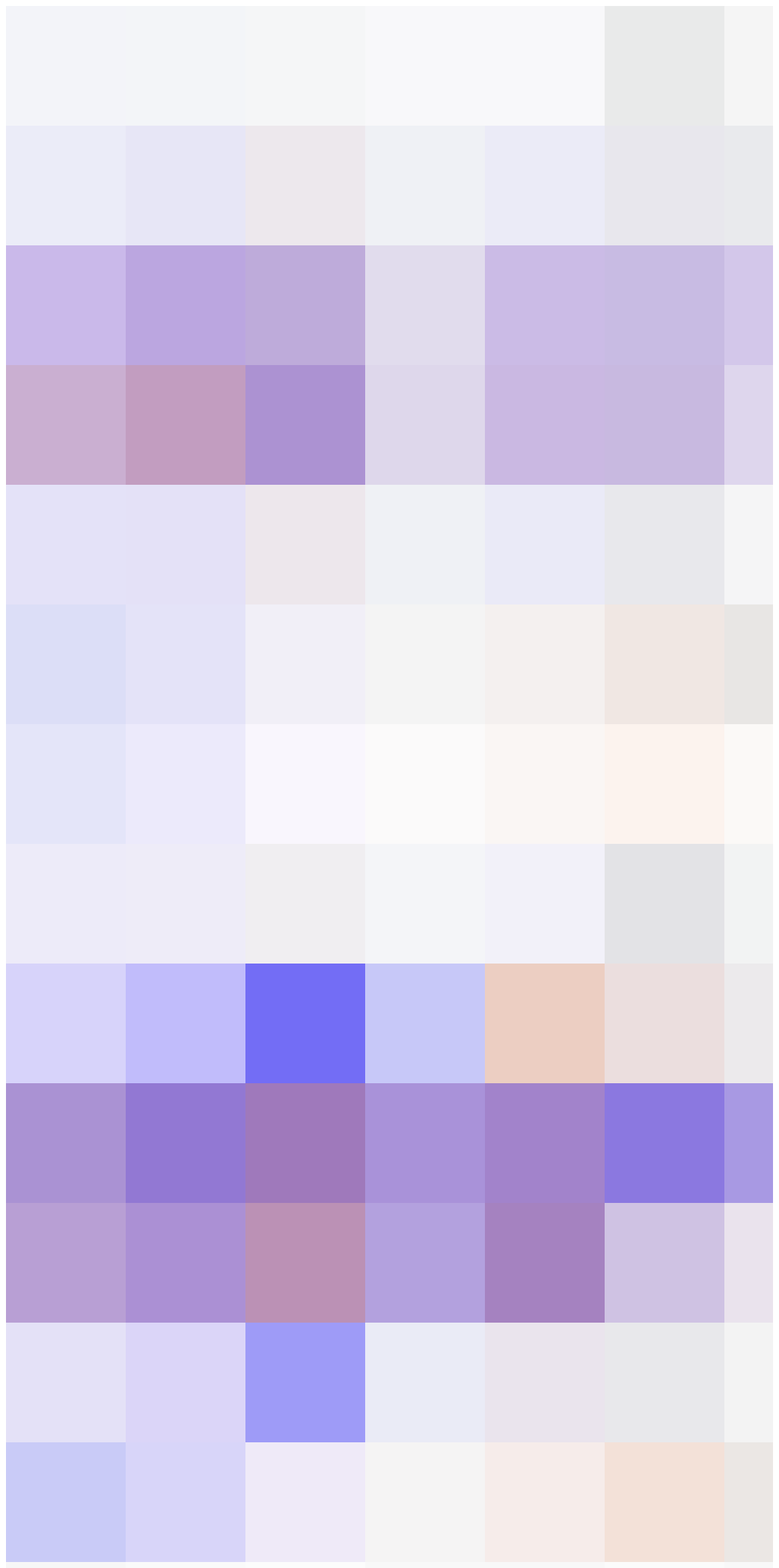}
\end{center}
\caption{The distribution of mean azimuthal magnetic fields 
         for model IV ($\beta_{b} = 1000$) at (a) $t = 824 t_{0} 
         ( = 3.1 \,\, \mathrm{Gyr})$ 
         and (b) $t = 1115 t_{0} ( = 4.2 \,\, \mathrm{Gyr})$.
         Blue and orange indicate regions where the mean field 
         $\bar{B_{y}}$ threading the $y=0$ plane is
         positive or negative, respectively.
         Arrows show the direction of mean magnetic fields.
         The mean azimuthal fields change their direction with height
         from the equatorial plane. 
        }
\label{fig:f18}
\end{figure}

We also carried out a simulation (model IV) starting from the initially weaker 
magnetic field ($\beta_{b} = 1000$). 
In this model, the magnetic energy saturates at almost
the same level as in model III ($B_{\varphi} \sim 1 \mu \,\, \mathrm{G}$ at
$\varpi \sim 5 \,\, \mathrm{kpc}$). 
Figure \ref{fig:f18} shows the spatial distribution of mean azimuthal
magnetic fields for model IV at $t = 824t_{0} (= 3.1 \,\, \mathrm{Gyr})$ and 
at $t = 1115t_{0} (= 4.2 \,\, \mathrm{Gyr})$. 
Blue and orange indicate regions where mean magnetic field 
$\bar{B_{\varphi}}$ threading the $y = 0$ plane is positive or 
negative, respectively. 
Again, the azimuthal magnetic field changes direction with radius and
with height. 
The striped distribution of azimuthal magnetic fields and the vertical
propagation of the reversals of azimuthal magnetic fields are similar to
those in model III (figure \ref{fig:f15}).

\begin{figure}[p!]
\begin{center}
\epsscale{0.7}
\plotone{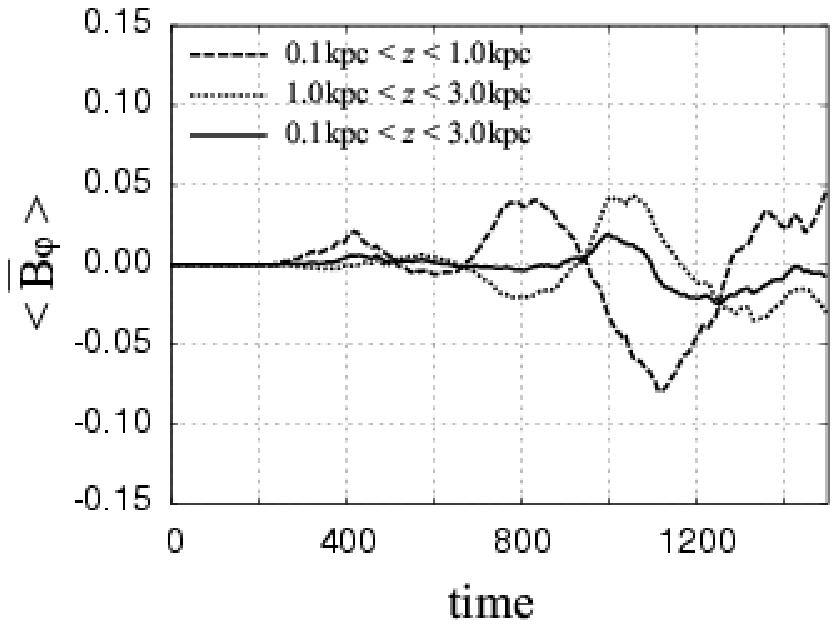}
\end{center}
\caption{The time variation of mean azimuthal field of model IV 
         averaged in $5 \,\, \mathrm{kpc} < \varpi < 6 \,\, \mathrm{kpc}$, 
         $0 \leq \varphi < 2 \pi$ and
         $0.1 \,\, \mathrm{kpc} < z < 1 \,\, \mathrm{kpc}$ (dashed curve), 
         $1.0 \,\, \mathrm{kpc} < z < 3.0 \,\, \mathrm{kpc}$ (dotted curve) and
         $0.1 \,\, \mathrm{kpc} < z < 3.0 \,\, \mathrm{kpc}$ (solid curve), respectively.
         The unit of $B_{\varphi}$ is $\sqrt{\rho_{b}v_{0}^{2}} = 15 \mu \,\, \mathrm{G}$
         in this model when $\rho_{b} = 3 \times 10^{-25} \,\, \mathrm{g/cm}^{3}$.
        }
\label{fig:f19}
\end{figure}

In Figure \ref{fig:f19} 
we plotted the time variation of mean azimuthal field of model IV. 
Magnetic fields are averaged in azimuthal direction 
($0 \leq \varphi <  2 \pi$), radial direction 
($5 \,\, \mathrm{kpc} \leq \varpi < 6 \,\, \mathrm{kpc}$), and 
in $0.1 \,\, \mathrm{kpc} < z < 1.0 \,\, \mathrm{kpc}$ (equatorial region), 
$1.0 \,\, \mathrm{kpc} < z < 3.0 \,\, \mathrm{kpc}$ (coronal region) and 
$0.1 \,\, \mathrm{kpc} < z < 3.0 \,\, \mathrm{kpc}$ (disk + corona), 
respectively. 
These results indicate that the direction of mean azimuthal magnetic
field reverses several times. 
The amplitude of oscillation of azimuthal magnetic fields increases with
time. 
The interval of reversal of azimuthal magnetic fields is $\sim 1
\,\, \mathrm{Gyr}$ (2 Gyr for a full cycle). 
This timescale coincides with that in model III, in which the initial
plasma $\beta$ is 100. 
The direction of azimuthal magnetic field in the coronal region is opposite 
to that of the equatorial region.

\subsection{Dependence on the Size of the Azimuthal Region and Mesh Size}

In order to study the dependence of numerical results on
the size of the azimuthal simulation region and mesh sizes in the
azimuthal direction, we carried out simulations in which we used 64
(model V), 32 (model VI) and 16 (model VII) grid points 
in $0 \leq \varphi < \pi/2$. 
Other parameters are the same as those in model III. 

\begin{figure}[p!]
\begin{center}
\epsscale{1.0}
\plotone{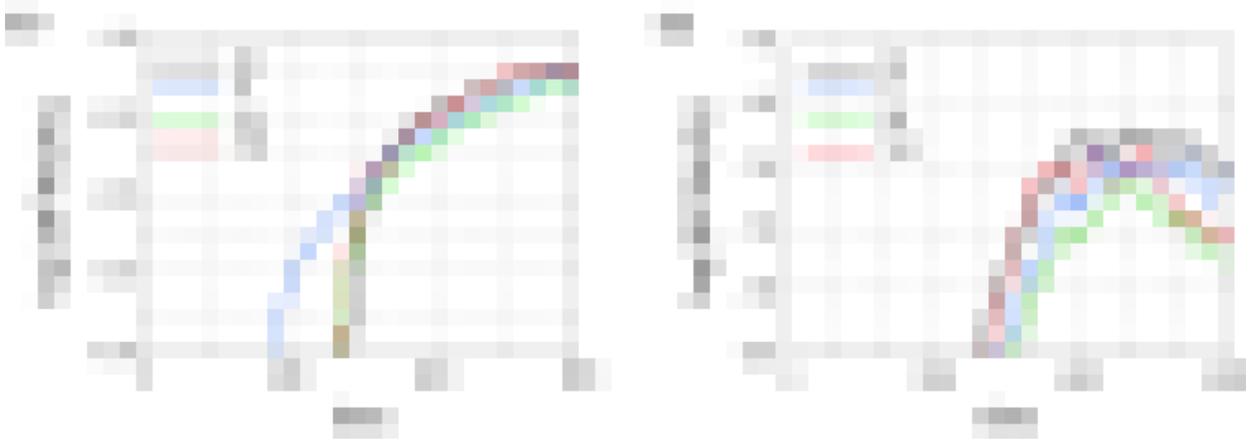}
\end{center}
\caption{The dependence of azimuthally averaged magnetic energy on the 
         azimuthal grid resolution and 
         the size of the azimuthal simulation region. 
         (a) Field amplification stage and (b) non-linear saturation stage. 
         Magnetic energy is averaged in $2 \,\, \mathrm{kpc} < \varpi < 5 \,\, \mathrm{kpc}$, 
         and $0 \,\, \mathrm{kpc} < z < 1 \,\, \mathrm{kpc}$. 
	 Azimuthal simulation region is $2\pi$ for model III and $\pi/2$ 
         for model V, VI and VII. 
	 The azimuthal grid size is $\pi/32$ for model III, and $\pi/128$,
	 $\pi/64$ and $\pi/32$ for model V, VI and VII, respectively.
        }
\label{fig:f20}
\end{figure}

Figure \ref{fig:f20}{\it a} shows the early stage ($t < 600t_{0}$)
of the magnetic field amplification in $2 \,\, \mathrm{kpc} < \varpi < 5
\,\, \mathrm{kpc}$ and figure \ref{fig:f20}{\it b} shows the time variation of
magnetic energy in the nonlinear saturation stage. 
We found that in the early stage ($t < 600t_{0}$), the growth of the magnetic
energy in model VII is similar to that of model III, which has the same
azimuthal grid resolution. 
In the nonlinear saturation stage (figure \ref{fig:f20}{\it b}), however, the
magnetic energy in model VII decreases and becomes smaller than that in
model III. 
These results are consistent with the results of local 3D MHD
simulations of accretion disks \citep*{Haw95}, in which they showed that
the nonlinear saturation level of the magnetic energy increases with the
size of the simulation region. 

When the azimuthal size of the simulation region is fixed (model V, VI and VII), 
the saturation level of the magnetic energy ($t > 600t_{0}$) is larger
in the high resolution model (model V) than that in low resolution
models (model VI and VII).
This result is also consistent with the results of local 3D MHD
simulations of the growth of MRI in accretion disks
\citep*[e.g.,][]{Haw95,Haw96}. 

The time evolution of the magnetic energy before its
saturation (figure \ref{fig:f20}{\it a}) depends on how the gas infalls and
carries in magnetic fields. 
The earlier rise of magnetic energy in the highest resolution model (model
V) is due to the faster growth of the MRI in the torus. 
The growth rate of the MRI is sensitive to the azimuthal grid resolution
because the most unstable wavenumber $k_{\parallel} \sim \Omega / v_{A}$, 
where $k_{\parallel}$ is the wavenumber parallel to the unperturbed
magnetic fields \citep*[e.g.,][]{Bal92}
is hardly resolved in low resolution simulations. 
When $\beta \sim 100$, the most unstable wavelength is 
$\lambda_{max} \sim 2\pi v_{A} / \Omega \sim 2\pi H(2/\beta)^{1/2} \sim H$ 
where $H$ is the disk half thickness. 
Thus, $\lambda_{max} \sim 0.5 \,\, \mathrm{kpc} \sim (\pi/2) \times 10
\,\, \mathrm{kpc} /30$. 
We need at least 120 mesh points in $0 < \varphi <
\pi/2$ when one wavelength is resolved by 4 mesh points. 
Thus, in all models we adopted in this paper, the fastest growing mode
is not numerically resolved. 
These estimates indicate that the growth rate of MRI is affected by the
grid size and that the growth rate in models III, VI and VII at $\varpi
= 10 \,\, \mathrm{kpc}$ are smaller than that in model V.  
Therefore, the infall of the torus gas is delayed in low-resolution models. 

Let us discuss why magnetic energy in low resolution
models (model III and VII) is larger than that in high resolution models
(model V and VI) in the nonlinear stage before the saturation ($t < 600t_{0}$). 
These magnetic fields are carried into $2 \,\, \mathrm{kpc} < \varpi < 5
\,\, \mathrm{kpc}$ with the gas infalling along the channels of magnetic fields. 
Since non-axisymmetric parasitic instabilities \citep*{Goo94} and
magnetic reconnection \citep*[e.g.,][]{San01,Mac03} which break up the
channel flow can be captured better in high resolution simulations, the growth
rate of the magnetic energy becomes smaller in high resolution models
than that in low resolution models. 

\begin{figure}[p!]
\begin{center}
\epsscale{1.0}
\plotone{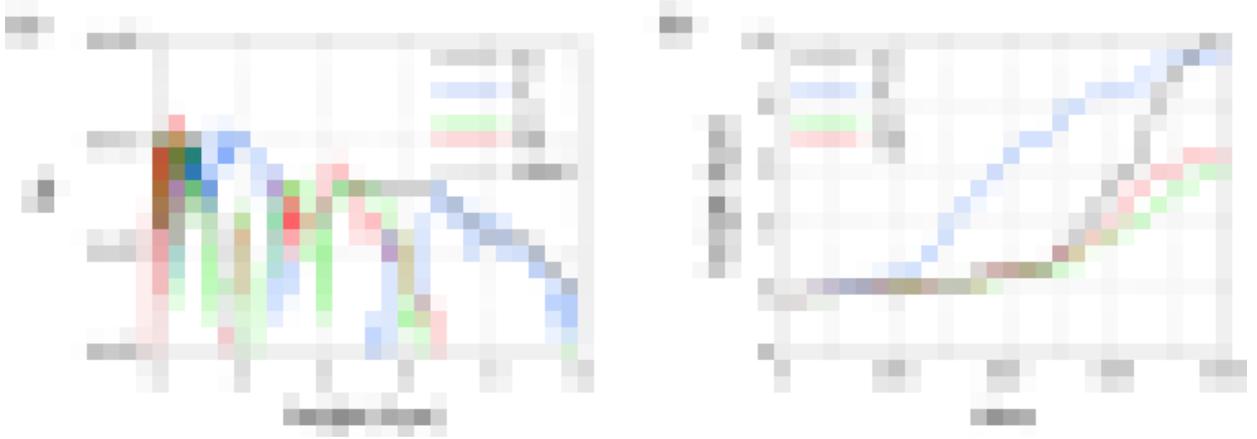}
\end{center}
\caption{(a) Vertical distribution of azimuthal magnetic field
         $B_{\varphi}$ at $\varpi = 10 \,\, \mathrm{kpc}$ 
         when $t = 800t_0$ for model III, V, VI and VII. 
         The dashed curve shows the initial profile of $B_{\varphi}$.
         (b) The time evolution of the rising magnetic flux for model
         III, V, VI and VII. 
         The height of the wave front of the rising magnetic flux where
         $B_{\varphi} = 0.003 \sqrt{\rho_{b} v_{0}^{2}}$ is plotted.  
        }
\label{fig:f21}
\end{figure}

Figure \ref{fig:f21}{\it a} compares the vertical distribution
of azimuthal magnetic fields at $\varpi = 10 \,\, \mathrm{kpc}$ at $t = 800t_{0}$
in model III, V, VI and VII. 
The region where $B_{\varphi} < 0$ is not plotted. 
The azimuthal magnetic flux buoyantly rises into the corona and forms
extended magnetized region. 
Numerical results indicate that the magnetic flux rises faster into the
corona in the high resolution model (model V) than low resolution
models (model VI and VII). 
High resolution simulations could resolve the most unstable wavelength
of the Parker instability, $\lambda \sim 10 H \sim 5 \,\, \mathrm{kpc}$ in
our simulation model \citep*[e.g.,][]{Par66,Mat88}.
Thus, one wavelength of the Parker unstable loop is well resolved in high
resolution model with 64 grid points in $0 \leq \varphi < \pi/2$ at 
$\varpi = 10 \,\, \mathrm{kpc}$ (model V) but hardly resolved in the low
resolution model (model VII). 
This may be the reason why toroidal magnetic flux rises faster in model V
than in other models (figure \ref{fig:f21}{\it b}). 

When the azimuthal resolution is fixed (model III and
VII), the magnetic flux rises faster in the model with larger azimuthal
simulation region (model III). 
This dependence on the azimuthal simulation region comes from the
horizontal expansion of the rising magnetic loops \cite[e.g.,][]{Shi89}. 
When the horizontal simulation region is restricted, since magnetic
tension prevents the escape of the magnetic flux, 
the speed of the buoyantly rising magnetic loops is reduced.

\subsection{Difference Between the Rotation Curve of Dark Matter and Gas}

\begin{figure}[p!]
\begin{center}
\epsscale{0.7}
\plotone{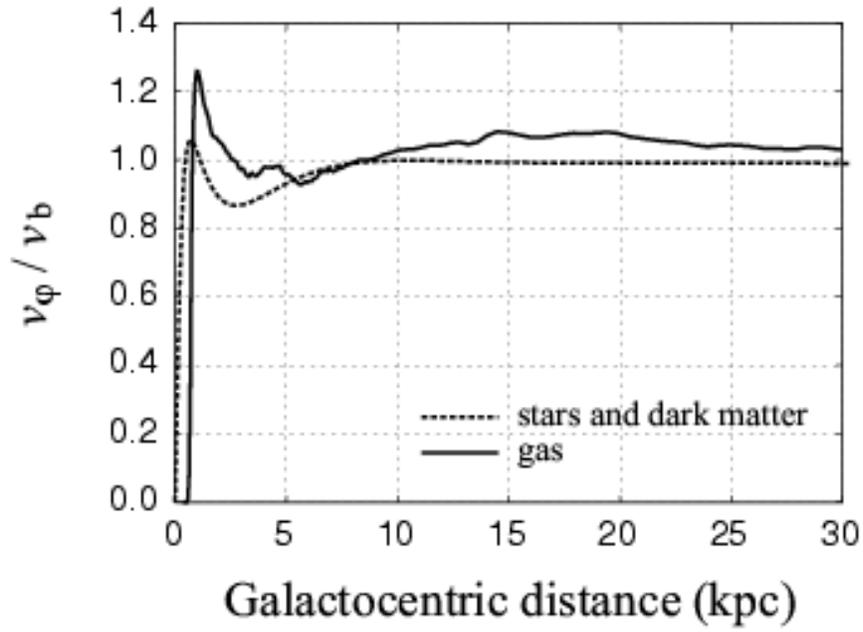}
\end{center}
\caption{The rotation curve of stars and dark matter computed from the 
         gravitational potential (dashed curve) and the rotation curve for 
         gas (solid curve) obtained by simulations for model III 
         at $t = 3.8 \,\, \mathrm{Gyrs}$. 
        }
\label{fig:f22}
\end{figure}

Figure \ref{fig:f22} compares the rotation curve of stars and dark matter computed 
from the gravitational potential (dashed curve) and the rotation curves for 
gas (solid curve) obtained from simulations for model III. 
The rotation curves of our Galaxy interior to the Sun
has been obtained by HI gas \citep*[e.g.,][]{Gun79}, and 
CO \citep*{Bur78} observations. 
The rotation speed in $5-10 \,\, \mathrm{kpc}$ from the galactic center is
$220-260 \,\, \mathrm{km/s}$ and similar to the rotation speed shown in 
figure \ref{fig:f22}. 
The rotation speed of our galaxy distant from the galactic center
($\varpi > 10 \,\, \mathrm{kpc}$) is obtained by radio observation of 
HI gas \citep*[e.g.,][]{Mer92}. 
The rotation speed is nearly flat and $v_{\varphi} \sim 240 \,\, \mathrm{km/s}$. 
The flat rotation curve is similar to the rotation curve in the outer
part of external spiral galaxies \citep*[e.g.,][]{Rub85}. 
Our numerical result reproduces these flat rotation curves when we use
the gravitational potential which includes the dark matter. 

Since the rotation curve of our galaxy distant from the galactic center is 
obtained by radio observations of HI gas, the observationally measured
rotation curve is that for the gas.
The rotation curve of the neutral hydrogen does not necessarily coincide with that 
of stars and the dark matter. 
Since they are almost frozen to the magnetic fields, 
if the magnetic arm rotates with angular speed different from the dark matter, 
the neutral hydrogen is not a good tracer of mass distribution in our Galaxy. 
\citet*{Bat92} suggested that when magnetic
pinch force is large enough, they can explain the observed rotation
curve of the Galaxy without dark matter. 
On the other hand, \citet*{San04} reported that the
contribution of the magnetic field to the rotation curve is up to 20
km/s and it is not sufficient to banish the effect of dark matter. 
Our numerical results indicate that even in the presence of dynamically 
non-negligible magnetic fields, gas 
component shows a flat rotation curve similar to that observed by neutral
hydrogen.
Numerical results for model I indicates that
if the dark matter does not exist, gas rotation curve decreases with radius.
Thus, the flat rotation curve obtained by observations of neutral hydrogen
really indicates the existence of dark matter.

\section{Summary and Discussion}

We carried out global three-dimensional magnetohydrodynamic (MHD) simulations 
of galactic gaseous disks to investigate how the galactic magnetic fields are 
amplified and maintained. 
We showed the dependence of numerical results on 
physical conditions in the innermost region, 
gravitational potential, and strength of initial magnetic fields. 

In all models, the magnetorotational instability (MRI) grows in the disk and 
the magnetic energy is amplified and maintained 
for more than $5 \,\, \mathrm{Gyrs}$. 
The saturation level of magnetic fields, however, is smaller than that 
expected from observations ($B \sim$ several $\mu G$) at $\varpi = 10 \,\, \mathrm{kpc}$. 
The multiphase nature of interstellar space, supernova explosions,
cosmic rays  and/or non-axisymmetric spiral potential may be important
to further amplify magnetic fields.

Numerical results indicate that mean azimuthal field inside the galactic disk 
is amplified and that its direction changes quasi-periodically. 
The amplitude of the oscillation of the azimuthal field increases with time. 
The interval of reversal of equatorial azimuthal field is $\sim 1 \,\, \mathrm{Gyr}$. 
This timescale is comparable to the timescale of the buoyant rise of
azimuthal magnetic flux from the disk to the corona. 
The amplitude of the fluctuating field is comparable to the 
mean magnetic field. 
The strength of the mean field only weakly depends on the initial strength 
of magnetic fields. 
In the coronal region, the mean azimuthal field has direction opposite to the 
mean field of the disk. 

Mean magnetic fields obtained from numerical simulations also 
show reversals in the radial direction near the equatorial plane around 
$\varpi = 5 \,\, \mathrm{kpc}$ at $t = 3.8 \,\, \mathrm{Gyr}$. 
The radius of the field reversal changes with time because the spiral
magnetic channels which produce the reversal of azimuthal fields in the
radial direction move inward as the disk gas accretes toward the
galactic center. 

\begin{figure}[p!]
\begin{center}
\epsscale{0.7}
\plotone{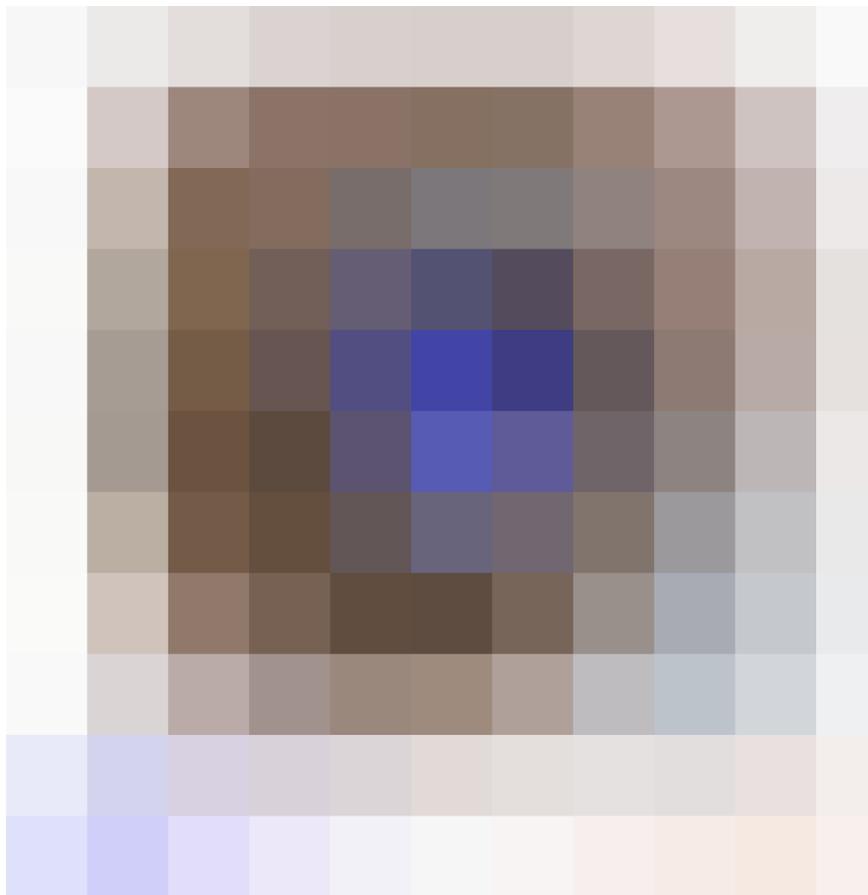}
\end{center}
\caption{Magnetic field lines depicted by mean magnetic
         fields (solid curves) and isocontours of azimuthal component of
         magnetic fields at $z = 0.25 \,\, \mathrm{kpc}$ (color) at 
         $t = 3.8 \,\, \mathrm{Gyr}$ for model III. 
         Regions colored in orange or blue show domains where azimuthal
         magnetic field at $z = 0.25 \,\, \mathrm{kpc}$ is positive or negative,
         respectively. 
         The box size is $30\,\, \mathrm{kpc} \times 30\,\, \mathrm{kpc}$.
        }
\label{fig:f13}
\end{figure}

In the Galactic plane, magnetic field reversals are
observed by Rotation Measure (RMs) of pulsars \citep*[e.g.,][]{Han02}. 
The reversals of mean magnetic fields are observed in our Galaxy between
the local Orion arm and the inner Sagittarius arm \citep*{Sim79}. 
More recent analysis imply axisymmetric field with two reversals 
(\citealt*{Ran89,Ran94}; see also \citealt*{Bec96}). 
Simulation results of model III (Figure \ref{fig:f13})
show reversals of mean magnetic fields near the equatorial plane around
$\varpi = 5 \,\, \mathrm{kpc}$. 
Note that RMs measure the mean magnetic field in some specific
direction from the Earth. 
When we draw lines from a point in Figure \ref{fig:f13}, we can see
field reversals in several directions. 
Since the magnetic fields inside the galactic disk is turbulent, more
and more field reversals will be observed as angular resolution of RMs
increase. 

In conventional theories of $\alpha \omega$-dynamos,
various kinds of spatial and temporal oscillations appear depending on
the dynamo number $D = C_{\alpha} C_{\Omega}$ where 
$C_\Omega = H^{2} \Omega/ \eta$ ($H$ is the thickness of the disk, $\Omega$ is 
the rotation angular speed and $\eta$ is the magnetic diffusivity) and 
$C_\alpha = H \alpha/\eta$ \citep*[e.g.,][]{Ste88}. 
In our global MHD simulations, we do not need to introduce the dynamo 
$\alpha$ parameter. 
The amplification of magnetic fields due to MRI and the reproduction of
toroidal fields through gas motions are self-consistently incorporated
in our simulations. 
We introduced anomalous resistivity $\eta$ to handle magnetic reconnection
which takes place in local region where current sheets are formed, but
the turbulent diffusivity is not explicitly assumed. 
We showed by direct 3D simulations that mean toroidal
magnetic fields show reversals both in space and time. 

\citet*{Tou92} constructed phenomenological models of disk
dynamos by taking into account the growth of magnetic fields due to MRI,
escape of magnetic flux by the Parker instability, and the dissipation
of magnetic fields by magnetic reconnection. 
They showed that the magnetic fields in
the disk oscillate quasi-periodically with period comparable to the time
scale of the buoyant escape of the magnetic flux. 
This timescale is consistent with the timescale of field reversals
obtained from our MHD simulations. 

Let us discuss the similarities and dissimilarities of
the magnetic activity of galactic gas disks produced by our numerical
simulations and those in the Sun. 
Both objects are differentially rotating plasma with high conductivity
in which magnetic fields are almost frozen to the rotating plasma.
Thus the magnetic fields can be deformed and amplified by the plasma
motion.

The Sun is a slow rotator in which the centrifugal force
is much smaller than the gravity, thus its atmosphere is confined by the
gravity.
The convective motions and/or the differential rotation in the
convection region may be the source of magnetic field amplification in
the Sun.
However, since the magnetic flux tube in the convection region buoyantly
rises with a timescale of the order of a month \citep[e. g., ][]{Par75},
much shorter than the solar magnetic cycle ($\sim 22 \,\, \mathrm{year}$),
solar dynamo activity is considered to be operating in the interface
between the convection zone and the radiative zone, where magnetic
buoyancy is small and differential rotation exists
\citep*[e. g., ][]{Spi80,Sch83}.
On the other hand, galactic gas disks are supported by the rotation and
subject to differential rotation, which induces the growth of MRI in the
whole disk. 
In galactic gas disks, since the gas disk is convectively stable, the
buoyant escape of the magnetic flux takes place with timescale 
$t_{buoy} \sim 10 H /v_{A} \sim 10 (1/\Omega)(\beta/2)^{1/2}$. 
This timescale is longer than the growth time of MRI $t_{MRI} \sim 1/\Omega$. 
Thus the magnetic fields can be amplified in the disk before escaping to
the corona. 

Numerical results indicate that the total azimuthal
magnetic flux is nearly conserved. 
When turbulent diffusivity exists, the azimuthal magnetic flux 
is not necessarily conserved. 
In our simulations, however, since magnetic diffusion
only exists in the local current sheet, the magnetic flux is almost
conserved. 
We found that the toroidal 
magnetic flux in the disk buoyantly rises into the corona. 
This result indicates that the buoyant escape of the magnetic flux from
the disk enables the amplification of mean magnetic fields in the
galactic gas disk. 
When the magnetic diffusion is not negligible, the total azimuthal
magnetic flux is not conserved but the magnetic helicity is conserved. 
\citet*{Bla03} pointed out the importance of helicity
conservation and the role of coronal mass ejections, which help
sustaining the solar dynamo cycle. 
In our simulations, buoyantly rising magnetic loops carry helicity as
well as the magnetic flux threading the disk. 
This escape of magnetic flux and helicity enable the disk to amplify the
azimuthal magnetic flux in the equatorial region.

We showed that when we remove the central absorber at $r = 0.8 \,\, \mathrm{kpc}$, 
formation of dense gas bulge forces the rotation curve to deviate from
that expected from the gravitational potential (model II). 
This indicates that in the central region of the galaxy, the gas absorption 
or the conversion of 
the accumulated gas to stars are essential to explain the observed rotation curve. 
The mass accretion rate to the central region is 
$\sim 10^{-3} M_{\odot}/ \,\, \mathrm{yr}$ 
when we adopt the absorbing boundary condition. 
This accretion rate depends on the initial density of the torus, which can 
be formed by the infall of intergalactic matter or by the supernova
explosions after bursts of star formation in the certain radius of the
disk.  
When the galaxy was more gas rich in the early stage of its evolution, 
the accretion rate should be much higher.

We also showed that the gas rotation curve approximately coincides with the 
rotation curve 
for stars and dark matter even when magnetic fields are dynamically important. 
This justifies us to use the gas rotation curve to estimate the distribution 
of the dark matter.

\acknowledgments

We thank Drs. T. Kuwabara, S. Miyaji and K. Shibata for discussion. 
Discussions at IAU XXXth General Assembly, 2003 and 
the workshop on Magnetic Fields in the Universe, 2004 helped us to complete 
this work. 
This work is supported by the Grants-in Aid  of Ministry of Education, Science, 
Sports and Culture (16340052, P.I., R. Matsumoto) the priority research 
project of the Graduate School of Science and Technology, Chiba
University, ACT-JST of the Japan Science and Technology corporation, 
and UK-Japan collaboration on magnetic activities of the Sun, 
Stars and accretion disks (P.I., K. Shibata and N. Weiss).
Numerical computations were carried out on VPP5000 at 
the Astronomical Data Analysis Center of the National Astronomical 
Observatory, Japan (NAOJ).

\clearpage
\newpage
\begin{center}
\begin{deluxetable}{lccccccccccc}
\tabletypesize{\footnotesize}
\tablewidth{0pt}
\tablecaption{Model parameters and inner boundary conditions}
\tablehead{\colhead{Models}&
           \colhead{$a_{1}$}&
           \colhead{$a_{2}$}&
           \colhead{$a_{3}$}&
           \colhead{$b_{1}$}&
	   \colhead{$b_{2}$}&
           \colhead{$b_{3}$}&
           \colhead{$M_{1}$}&
	   \colhead{$M_{2}$}&
           \colhead{$M_{3}$}&
           \colhead{$\beta_{b}$}&
           \colhead{Boundary}
          }
\startdata
Model I   &$0.0$&$7.258$&$0.0$&$0.495$&$0.520$&$0.0$&$2.05$&$25.47$&$0.0$&100&Absorbing\\
Model II  &$0.0$&$7.258$&$0.0$&$0.495$&$0.520$&$0.0$&$2.05$&$25.47$&$0.0$&100&Non-Absorbing\\
Model III &$0.0$&$6.2$&$0.0$&$0.47$&$0.15$&$31.2$&$1.95$&$17.4$&$73.5$&100&Absorbing\\
Model IV  &$0.0$&$6.2$&$0.0$&$0.47$&$0.15$&$31.2$&$1.95$&$17.4$&$73.5$&1000&Absorbing\\
Model V-VII &$0.0$&$6.2$&$0.0$&$0.47$&$0.15$&$31.2$&$1.95$&$17.4$&$73.5$&100&Absorbing\\
\enddata
\label{tbl:tbl1}
\end{deluxetable}
\end{center}

\begin{center}
\begin{deluxetable}{llcc}
\tablewidth{0pt}
\tablecaption{Azimuthal grid size, number of grid points in azimuthal direction 
              ($N_{\varphi}$), and the azimuthal size of the simulation region}
\tablehead{\colhead{Models}&
	   \colhead{$\Delta \varphi$}&
           \colhead{$N_{\varphi}$}&
           \colhead{simulation region}
          }
\startdata

Model I-IV   &$\pi/32$  &64 &$0 \leq \varphi < 2\pi$ \hspace{3pt} \\
Model V      &$\pi/128$ &64 &$0 \leq \varphi < \pi/2$\\
Model VI     &$\pi/64$  &32 &$0 \leq \varphi < \pi/2$\\
Model VII    &$\pi/32$  &16 &$0 \leq \varphi < \pi/2$\\
\enddata
\tablecomments{Other parameters of model V, VI and VII are 
               the same as those in model III.}
\label{tbl:tbl2}
\end{deluxetable}
\end{center}

\clearpage
\newpage

\clearpage

\begin{thebibliography}{}
\bibitem[Balbus et al.(1991) Balbus \& Hawley]{Bal91} 
         Balbus, S. A., \& Hawley, J. F.
    1991, \apj, 376, 214
\bibitem[Balbus et al.(1992) Balbus \& Hawley]{Bal92} 
         Balbus, S. A., \& Hawley, J. F.
    1992, \apj, 400, 610
\bibitem[Battaner et al.(1992)]{Bat92} 
         Battaner, E., Garrido, J. L., Membrado, M., \& Florido, E. 
    1992, \nat, 360, 652
\bibitem[Beck et al.(1996)]{Bec96} 
         Beck, R., Brandenburg, A., Moss, D., Shukurov, A., \& Sokoloff, D. 
    1996, \araa, 34, 155
\bibitem[Blackman et al.(2003) Blackman \& Brandenburg]{Bla03} 
         Blackman, E. G., \& Brandenburg, A.
    2003, \apjl, 584, 99
\bibitem[Brandenburg et al.(1989)]{Bra89} 
         Brandenburg, A., Krause, F. Meinel, R., Moss, D., \& Tuominen, I.
    1989, \aap, 213, 411
\bibitem[Brandenburg et al.(1992)]{Bra92} 
         Brandenburg, A., Donner, K. J.,Moss, D., Shukurov, A., 
         Sokolov, D. D., \& Tuominen, I.
    1992, \aap, 259, 453
\bibitem[Brandenburg et al.(1995)]{Bra95} 
         Brandenburg, A., Nordlund, A., Stein, R. F., \& Torkelsson, U.
    1995, \apj, 446, 741
\bibitem[Burton et al.(1978) Burton \& Gordon]{Bur78} 
         Burton, W. B., \& Gordon, M. A.
    1978, \aap, 63, 7 
\bibitem[Dziourkevitch et al.(2003) Dziourkevitch \& Elstner]{Dzi03} 
         Dziourkevitch, N., \& Elstner, D. 
    2003, \apss, 284, 757
\bibitem[Dziourkevitch et al.(2004) Dziourkevitch, Elstner \& R\"{u}diger]{Dzi04} 
         Dziourkevitch, N., Elstner, D., \& R\"{u}diger, G. 
    2004, \aap, 423, L29
\bibitem[Foglizzo et al.(1994) Foglizzo \& Tagger]{Fog94} 
         Foglizzo, T., \& Tagger, M. 
    1994, \aap, 287, 297
\bibitem[Foglizzo et al.(1995) Foglizzo \& Tagger]{Fog95} 
         Foglizzo, T., \& Tagger, M. 
    1995, \aap, 301, 293
\bibitem[Goodman et al.(1994) Goodman \& Xu]{Goo94} 
         Goodman, J.,\& Xu, G.
    1994, \apj, 432, 213 
\bibitem[Gunn et al.(1979) Gunn, Knapp \& Tremaine]{Gun79} 
         Gunn, J. E., Knapp, G. R., \& Tremaine, S. D.
    1979, \aj, 84, 1181
\bibitem[Han et al.(2002)]{Han02} 
         Han, J. L., Manchester, R. N., Lyne, A. G., \& Qiao, G. J.
    2002, \apjl, 570, L17 
\bibitem[Hawley et al.(1992) Hawley \& Balbus]{Haw92} 
         Hawley, J. F., \& Balbus, S. A.
    1992, \apj, 400, 595 
\bibitem[Hawley et al.(1995) Hawley, Gammie \& Balbus]{Haw95} 
         Hawley, J. F., Gammie, C. F., \& Balbus, S. A.
    1995, \apj, 440, 742
\bibitem[Hawley et al.(1996) Hawley, Gammie \& Balbus]{Haw96} 
         Hawley, J. F., Gammie, C. F., \& Balbus, S. A.
    1996, \apj, 464, 690
\bibitem[Hawley(2000)]{Haw00} 
         Hawley, J. F. 
    2000, \apj, 528, 462
\bibitem[Hayashi et al.(1996) Hayashi, Shibata \& Matsumoto]{Hay96} 
         Hayashi, M. R., Shibata, K., \& Matsumoto, R.
    1996, \apjl, 468L, 37 
\bibitem[Kato et al.(2004) Kato, Mineshige \& Shibata]{Kat04} 
         Kato, Y., Mineshige, S., \& Shibata, K. 
    2004, \apj, 605, 307
\bibitem[Kim et al.(2003) Kim, Ostriker \& Stone]{Kim03} 
         Kim, W. T., Ostriker, E. C., \& Stone, J. M. 
    2003, \apj, 599, 1157
\bibitem[Kitchatinov et al.(2004) Kitchatinov, \& R\"{u}diger]{Kit04} 
         Kitchatinov, L. L., \& R\"{u}diger, G.
    2004, \aap, 424, 565
\bibitem[Machida et al.(2000) Machida, Hayashi \& Matsumoto]{Mac00} 
         Machida, M., Hayashi, M. R., \& Matsumoto, R.
    2000, \apj, 532, L67
\bibitem[Machida et al.(2003) Machida \& Matsumoto]{Mac03} 
         Machida, M., \& Matsumoto, R. 
    2003, \apj, 585, 429
\bibitem[Matsumoto et al.(1988)]{Mat88} 
         Matsumoto, R., Horiuchi, T., Shibata, K., \& Hanawa, T.
    1988, \pasj, 40, 171
\bibitem[Matsumoto et al.(1995) Matsumoto \& Tajima]{Mat95} 
         Matsumoto, R., \& Tajima, T.
    1995, \apj, 445, 767
\bibitem[Matsumoto et al.(1996)]{Mat96} Matsumoto, R., Uchida, Y., 
         Hirose, S., Shibata, K., Hayashi, M. R., Ferrari, A., Bodo, G., \& Norman, C.
    1996, \apj, 461, 115
\bibitem[Matsumoto(1999)]{Mat99} 
         Matsumoto, R.
    1999, in Numerical Astrophysics, ed. Miyama, S. M., Tomisaka, K., \& 
         Hanawa, T. (Amsterdam: Kluwer Academic Publishers) 195 
\bibitem[Merrifield(1992)]{Mer92} 
         Merrifield, M. R.
    1992, \aj, 103, 1552
\bibitem[Miller et al.(2000) Miller \& Stone]{Mil00} 
         Miller, K. A., \& Stone, J. M.
    2000, \apj, 534, 398
\bibitem[Miyamoto et al.(1975) Miyamoto \& Nagai]{Miy75} 
         Miyamoto, M., \& Nagai, R. 
    1975, \pasj, 27, 533
\bibitem[Miyamoto et al.(1980) Miyamoto, Satoh \& Ohashi]{Miy80} 
         Miyamoto, M., Satoh, C., \& Ohashi, M. 
    1980, \aap, 90, 215
\bibitem[Okada et al.(1989) Okada, Fukue \& Matsumoto]{Oka89} 
         Okada, R., Fukue, J., \& Matsumoto, R.
    1989, \pasj, 41, 133
\bibitem[Parker(1966)]{Par66} 
         Parker, E. N.
    1966, \apj, 145, 811
\bibitem[Parker(1971)]{Par71} 
         Parker, E. N.
    1971, \apj, 163, 255
\bibitem[Parker(1975)]{Par75} 
         Parker, E. N. 
    1975, \apj, 198, 205
\bibitem[Piontek et al.(2004) Piontek \& Ostriker]{Pio04} 
         Piontek, R. A., \& Ostriker, E. C. 
    2004, \apj, 601, 905
\bibitem[Rand et al.(1989) Rand \& Kulkarni]{Ran89} 
         Rand, R. J., \& Kulkarni, S. R.
    1989, \baas, 21, 1188
\bibitem[Rand et al.(1994) Rand \& Lyne]{Ran94} 
         Rand, R. J., \& Lyne, A. G.
    1994, \mnras, 268, 497
\bibitem[Richtmyer et al.(1967) Richtmyer \& Morton]{Ric67} 
         Richtmyer, R., O., \& Morton, K., W. 
    1967, Differential Methods for Initial Value Probrem (2d ed., New York: Wiley)
\bibitem[Rubin et al.(1967) Rubin \& Burstein]{Rub67} 
         Rubin, E., \& Burstein, S., Z. 
    1967, J. Comput. Phys., 2, 178
\bibitem[Rubin et al.(1985)]{Rub85} 
         Rubin, V. C., Burstein, D., Ford, W. K., \& Thonnard, N. 
    1985, \apj, 289, 81
\bibitem[Sanchez et al.(2004) S\'{a}nchez-Salcedo \& Reyes-Ruiz]{San04} 
         S\'{a}nchez-Salcedo, F. J., \& Reyes-Ruiz, M.     
    2004, \apj, 607, 247
\bibitem[Sano et al.(2001) Sano \& Inutsuka]{San01} 
         Sano, T., \& Inutsuka, S.
    2001, \apjl, 561, 179
\bibitem[Schmitt et al.(1989) Schmitt \& Sch\"{u}ssler]{Sch89} 
         Schmitt, D., \& Sch\"{u}ssler M.,
    1989, \aap, 223, 343
\bibitem[Schmitt et al.(1983) Schmitt \& Rosner]{Sch83} 
         Schmitt, J. H. M. M., \& Rosner, R. 
    1983, \apj, 265, 901 
\bibitem[Sellwood et al.(1999) Sellwood \& Balbus]{Sel99} 
         Sellwood, J. A., \& Balbus, S. A.,
    1999, \apj, 511, 660
\bibitem[Shibata et al.(1985) Shibata \& Uchida]{Shi85} 
         Shibata, K., \& Uchida, Y.
    1985, \pasj, 37, 31
\bibitem[Shibata et al.(1989) Shibata, Tajima, Steinolfson \& Matsumoto]{Shi89} 
         Shibata, K., Tajima, T., Steinolfson, R. S. \& Matsumoto, R.
    1989, \apj, 345, 584
\bibitem[Simard-Normandin et al.(1979) Simard-Normandin \& Kronberg]{Sim79} 
         Simard-Normandin, M., \& Kronberg, P. P.
    1979, \nat, 279, 115
\bibitem[Sofue et al.(1986) Sofue, Fujimoto \& Wielebinski]{Sof86} 
         Sofue, Y., Fujimoto, M., \& Wielebinski, R. 
    1986, \araa, 24, 459
\bibitem[Spiegel et al.(1980) Spiegel \& Weiss]{Spi80} 
         Spiegel, E. A., \& Weiss, N. O. 
    1980, \nat, 287, 616
\bibitem[Stepinski et al.(1988) Stepinski \& Levy]{Ste88} 
         Stepinski, T. F., \& Levy, E. H.
    1988, \apj, 331, 416
\bibitem[Tout et al.(1992) Tout \& Pringle]{Tou92} 
         Tout, C. A., \& Pringle, J. E.
    1992, \mnras, 259, 604
\bibitem[Yokoyama et al.(1994) Yokoyama \& Shibata]{Yok94} 
         Yokoyama, T., \& Shibata, K.
    1994, \apjl, 436, L197
\end{thebibliography}
\end{document}